\documentclass{jpp}
\usepackage{graphicx}
\usepackage{epstopdf, epsfig}

 \usepackage{amssymb}
 \usepackage{bigints}
 \usepackage{amsmath}
 \usepackage{empheq}
 \usepackage{framed}
 \usepackage{color}
 \usepackage{enumerate}
 \usepackage{ulem}
 \usepackage{bm}
 \usepackage{appendix}
 \usepackage{multicol}
 \numberwithin{equation}{section}
 \usepackage{upgreek}
 \usepackage{textgreek}
 \usepackage{graphicx}
\usepackage{stmaryrd} 
\usepackage{xfrac}    

\newcommand{\comment}[1]{}

\newcommand{\be}{\begin{equation}}
\newcommand{\ee}{\end{equation}}
\newcommand{\ba}{\[\begin{aligned}}
\newcommand{\ea}{\end{aligned}\]}
\newcommand{\bea}{\begin{eqnarray}}
\newcommand{\eea}{\end{eqnarray}}
\newcommand{\beann}{\begin{eqnarray*}}
\newcommand{\eeann}{\end{eqnarray*}}
\newcommand{\bs}{\begin{split}}
\newcommand{\es}{\end{split}}

\newcommand*{\cA}{\mathcal{A}} 

\newcommand*{\cI}{\mathcal{I}}

\newcommand*{\cL}{\mathcal{L}}

\newcommand*{\mfa}{\mathfrak{a}}

\newcommand*{\mff}{\mathfrak{f}}

\newcommand*{\ep}{\epsilon}

\newcommand*{\B}{\bm{B}}

\newcommand*{\J}{\bm{J}}

\newcommand*{\w}{\bm{w}}

\newcommand*{\BDB}{\B\cdot \dl B}

\newcommand*{\uD}{\u\cdot \dl }

\renewcommand*{\u}{\bm{u}}
\renewcommand*{\v}{\bm{v}}

\newcommand*{\dpl}{\nabla_{||}}

\newcommand*{\jpl}{j_{||}}

\newcommand*{\dl}{\bm{\nabla}}
\newcommand*{\del}{\partial}
\newcommand*{\BD}{\bm{B}\cdot\bm{\nabla}}

\newcommand*{\dlr}{\bm{\nabla_\perp}}
\newcommand*{\dlrs}{\nabla_\perp^2}
\newcommand*{\dlts}{\Delta^*}

\newcommand*{\lbr}{\left(}
\newcommand*{\rbr}{\right)}

\newcommand*\at[2]{\left.#1\right|_{#2}}

\newcommand*\reallywidetilde[1]{\ThisStyle{%
  \setbox0=\hbox{$\SavedStyle#1$}%
  \stackengine{-.1\LMpt}{$\SavedStyle#1$}{%
    \stretchto{\scaleto{\SavedStyle\mkern.2mu\sim}{.5467\wd0}}{.7\ht0}%
  }{O}{c}{F}{T}{S}%
}}

\def\test#1{$%
  \reallywidetilde{#1}\,
  \scriptstyle\reallywidetilde{#1}\,
  \scriptscriptstyle\reallywidetilde{#1}
$\par}
\makeatletter
\newsavebox{\@brx}
\newcommand{\llangle}[1][]{\savebox{\@brx}{\(\m@th{#1\langle}\)}%
  \mathopen{\copy\@brx\mkern2mu\kern-0.9\wd\@brx\usebox{\@brx}}}
\newcommand{\rrangle}[1][]{\savebox{\@brx}{\(\m@th{#1\rangle}\)}%
  \mathclose{\copy\@brx\mkern2mu\kern-0.9\wd\@brx\usebox{\@brx}}}
\makeatother

\shorttitle{QS-HBS} 

\title{Asymptotic quasisymmetric high-beta 3D MHD equilibria near axisymmetry}

\shortauthor{W. Sengupta, N. Nikulsin, R. Gaur, A. Bhattacharjee}
\author{Wrick Sengupta\aff{1,2}\corresp{\email{wsengupta@princeton.edu}}, Nikita Nikulsin \aff{1,2}, Rahul Gaur\aff{3}, Amitava Bhattacharjee\aff{1,2}}

\affiliation{
\aff{1} Department of Astrophysical Sciences, Princeton University, Princeton, NJ 08543, USA
\aff{2} Princeton Plasma Physics Laboratory, Princeton, NJ, 08540, USA
\aff{3}Department of Mechanical and Aerospace Engineering, Princeton University, Princeton, NJ 08544, USA
}

\begin{document}

\maketitle

\begin{abstract}
Quasisymmetry (QS), a hidden symmetry of the magnetic field strength, is known to support nested flux surfaces and provide superior particle confinement in stellarators. In this work, we study the ideal MHD equilibrium and stability of high-beta plasma in a large aspect-ratio stellarator. In particular, we show that the lowest-order description of a near-axisymmetric equilibrium vastly simplifies the problem of 3D quasisymmetric MHD equilibria, which can be reduced to a standard elliptic Grad-Shafranov equation for the flux function. We show that any large aspect-ratio tokamak, deformed periodically in the vertical direction, is a stellarator with approximate volumetric QS. We discuss exact analytical solutions and numerical benchmarks. Finally, we discuss the ideal ballooning and interchange stability of some of our equilibrium configurations.
 
\end{abstract}

\section{Introduction \label{sec:intro}}
An essential requirement of magnetic confinement is that the plasma must be in a stable magnetohydrodynamic (MHD) equilibrium. Since the confining magnetic field in a stellarator is mainly produced by external current-carrying coils or magnets, it does not suffer from the current-driven instabilities intrinsic to tokamaks. For a stellarator, a set of nested toroidal flux surfaces with a common magnetic axis such that the magnetic field lines are tangential to the surfaces is desired. However, the lack of continuous symmetry in a stellarator can lead to a breakdown of nested flux surfaces and the formation of island chains and ergodic field line regions. Without symmetry, the trapped particles can drift out radially and be lost. Both of these effects are severely detrimental to confinement. Furthermore, the lack of symmetry and the nonlinear nature of ideal MHD equations seriously complicate our understanding of stellarator equilibrium. Consequently, compared to tokamaks, MHD equilibrium theory for generic stellarators still has major unanswered questions.

A hidden symmetry of the magnetic field strength, called quasisymmetry (QS), can ameliorate some of the difficulties mentioned above. While the magnetic field remains fully three-dimensional (3D), QS requires that the field strength, $B$, be independent of one of the angular coordinates. The continuous symmetry of  $B$ then leads to a  conserved canonical angular momentum (due to Noether's theorem), ensuring neoclassical particle confinement of the same quality as that of a tokamak \citep{helander2014, landreman_catto_2010effects_of_radial_E}. Moreover, exact QS \footnote{\footnotesize{In the literature, a distinction is made between the so-called strong and weak forms of QS. However, the two forms are equivalent \citep{burby2020,constantin2020,rodriguez2020a} for the ideal MHD force balance considered here.}} leads to nested  flux surfaces (\cite{burby2020}, \cite{rodriguez2020a}). Indeed, one of the benefits of understanding the space of near-axisymmetric QS is that it provides a valuable perspective for 3D error-field correction of tokamaks \citep{Park_QA}, which are known to be sensitive to 3D perturbations \citep{Park_Boozer2007_IPEC,park_Menard_boozer2009importance}. If the 3D perturbations of an axisymmetric tokamak restore QS, they can help mitigate harmful transport effects \citep{park2007control_PRL,park20183d_Nature}. 

Unfortunately, exact QS is difficult, if not impossible, to obtain. Analytical results \citep{garrenboozer1991a,garrenboozer1991b,plunk2018,landreman2019,jorge2020construction} seem to indicate that imposing QS on ideal MHD with scalar pressure leads in general to an overdetermined problem, with no global (or volumetric) solution. In particular, formal series expansions carried out in the distance from the axis, the so-called near-axis expansion (NAE), show that beyond the second-order, there are more equations than unknowns. Large-scale numerical optimization approach has successfully provided multiple practical designs \citep{Anderson1995,Zarnstorff2001,Najmabadi2008,Ku2010,bader2019, landreman2021}. However, the many degrees of freedom arising from the three-dimensionality of stellarators lead to computational challenges in optimizing them. A well-known problem for multi-parameter optimization is getting stuck in local minima in parameter space, and the deviations from exact QS cannot generally be made arbitrarily small.

Significant analytical and numerical progress has been made in recent times in understanding the QS constraint, which has been achieved with very high precision \citep{landreman2021}. NAEs have provided helpful analytical insights into new previously unknown configurations and provided excellent initial guesses for further numerical optimization. It has also been possible to map out the entire quasisymmetric phase-space using second-order NAEs. The important question of how the magnetic field strength shapes magnetic flux surfaces can be addressed within the NAE framework. For a local equilibrium, one is free to prescribe the flux-surface shape and the magnetic field strength \citep{boozer2002local_eqb,skovoroda2007relationship}. However, for a global quasisymmetric equilibrium, the flux-surface shapes are highly constrained \citep{jorge2020construction,rodriguez2022_thesis}. Second-order NAE theory allows one to understand and explore this relationship \citep{rodriguez2023constructing,rodriguez2022_thesis}.

Despite the several advantages of NAE, a significant drawback, when applied to QS, is that physical quantities can be approximated only by low-order polynomials in the radial variable. The reason lies in the overdetermined nature of QS. Therefore, only plasma profile functions, such as the pressure, magnetic shear, and bootstrap current that can be sufficiently well described by low-order polynomials, can be treated within the NAE framework for QS. A more serious drawback is that the overdetermination problem (discussed above) becomes an obstacle to calculating global magnetic shear, which shows up at the order at which the NAE approach breaks down. In the absence of reliable information on the global magnetic shear, MHD stability analyses, such as those pertaining to ballooning or interchange modes, become unreliable. Therefore, an alternative to the NAE is needed for the equilibrium and stability analysis of quasisymmetric devices.

In this work, we present an alternative approach to QS, which provides analytical insight akin to NAE but allows us to compute approximate global equilibrium solutions with more general profile functions. The expansion parameter in our reduced MHD model is the ratio of the equilibrium perpendicular and parallel length scales ($L_\perp/L_\parallel$) to the lowest-order magnetic field. We order plasma beta to be of $O(L_\perp/L_\parallel)$. We further restrict the lowest-order magnetic field to be a purely toroidal vacuum field. With this restriction, we can only treat quasi-axisymmetric configurations. \cite{plunk2018} have shown that exact volumetric vacuum QS cannot be obtained close to vacuum axisymmetry under quite general conditions. Without any contradiction with \cite{plunk2018}, we are able to realize approximate volumetric QS in our model since we satisfy the ideal MHD force balance and QS only to the lowest nontrivial orders. We are also consistent with the CDG (Constantin-Drivas-Ginzberg) theorem \citep{constantin2021}, which states that approximate volumetric QS can be obtained if the force-balance condition in ideal MHD is modified by the addition of a small force, allowing one to satisfy two of the weak QS conditions exactly. We have imposed the near-axisymmetry restriction to leverage the simplicity of the lowest-order axisymmetric geometry but shall relax this restriction in a future study.

The reduced MHD structure of our model coincides with the traditional large aspect-ratio expansion (LAE) of ideal MHD \citep{hazeltine_meiss2003plasma_confinement_book}. An LAE model for stellarators, accurate to the first order in the inverse aspect ratio, with the plasma beta ordered as the inverse aspect ratio, is known in the literature \citep{freidberg2014idealMHD,wakatani1998stellarator_heliotron} as the High-Beta Stellarator (HBS) model. (The rotational transform is finite in this model.) Unlike earlier large-aspect-ratio stellarator models \citep{Greene_Johnson_1961_Stallarator_expansion,strauss1980stellarator_EOM,strauss_Monticello_1981limiting}, the HBS does not require a large toroidal periodicity mode number, $N$. Given the high interest in quasisymmetric stellarators, we impose the QS constraint on the HBS and assess some of the important consequences for MHD equilibrium and stability. 

The structure of the paper is as follows. In Section \ref{sec:QS_iMHD_basics}, after discussing the implications of QS for ideal MHD equilibria, we provide a derivation of the  Quasisymmetric Grad-Shafranov equation (QS-GSE) in reduced MHD. We then discuss the Quasisymmetric HBS (QS-HBS) model in Section \ref{sec:QS-HBS_model}, and in particular, we derive a special form of the QS-GSE. We present an analytic solution to the QS-GSE and its numerical verification in Section \ref{sec:Extended_soloviev}. We discuss ballooning and interchange stability of the analytical equilibria in Section \ref{sec:stability}. We conclude with a discussion of the implications of our results and possible generalizations in Section \ref{sec:discussion}.

\section{Ideal MHD with the quasisymmetry constraint }\label{sec:QS_iMHD_basics}



We begin with the well-known model for a plasma equilibrium, the ideal MHD equations,
\begin{subequations}
\begin{align}
    \dl \cdot \B=0 \label{divB},\\
     \J\times \B = \dl p  \label{JxB},\\ 
     \J=\dl \times \B.  \label{JiscurlB}
\end{align}
    \label{eq:ideal_MHD}
\end{subequations}
Here, $\B,\J$, and $p$ denote the magnetic field vector, the current density, and the plasma pressure, respectively. We have used units such that $\mu_0=1$. We will assume that pressure is constant on a set of nested toroidal flux surfaces labeled by $\psi$, i.e.,
\begin{align}
    p=p(\psi), \quad \BD\psi=0.
    \label{eq:p_of_psi}
\end{align}
We shall further assume that $\nabla \psi$ is nonzero almost everywhere except on a finite set of closed field lines that include the magnetic axis and the separatrix.

The QS constraint can be imposed on the ideal MHD equilibrium in many different but equivalent ways. Here, we shall make use of the following form, which is most commonly referred to as the two-term form \citep{rodriguez2020a, helander2014, burby2020}:
\begin{align}
   \BDB= \frac{1}{F(\psi)}\B \times \dl \psi \cdot \dl B 
    \label{eq:2term_QS}.
\end{align}
The flux-function $\psi$ denotes the toroidal flux, $B$ is the magnetic field strength and $F(\psi)$ is a flux-function that can be shown \citep{helander2014}  to be related to the rotational transform $\iota(\psi)$, through
\begin{align}
   \frac{1} {F(\psi)}=\frac{\iota(\psi)-N/M}{G(\psi)+ (N/M)I(\psi)}.
\end{align}
Here, $N/M$ is the helicity, and $G, I$ are poloidal and toroidal currents. For quasiaxisymmetry (QA), which is the subject of this work, $N=0$. The currents can be obtained by integrating $\B$ along the constant $\vartheta$ (poloidal) and constant $\varphi$ (toroidal) angles. These angles can be chosen as the Boozer angles for convenience. Thus,
\begin{align}
    G(\psi)=\frac{1}{2\pi}\oint_\vartheta \B\cdot \bm{dr}, \quad I(\psi)=\frac{1}{2\pi}\oint_\varphi \B\cdot \bm{dr}
    \label{eq:Boozer_G_and_I}.
\end{align}
 Another relevant quantity of interest is the quasisymmetry vector $\u$ \citep{rodriguez2020a,burby2020}, which may be defined by the following two equivalent relations:
\begin{align}
    \u = \frac{\dl \psi\times \dl B}{\BDB} ,\quad \u = \frac{1}{B^2}\lbr F(\psi)\B -\B \times \dl \psi \rbr
    \label{eq:QS_u}.
\end{align}
We note here that the QS vector of Burby et al. differs by a factor of iota in the quasi-axisymmetric case assumed here.
Therefore, the 2-term QS relation is equivalent to $\u\cdot \B =F(\psi)$. The QS vector $\u$ defines curves that lie on constant flux surfaces along which $B$ does not change since $\u\cdot \dl B=0$ and $\u\cdot \dl \psi=0$. Furthermore, $\u$ lines are closed since $B$ is single-valued. 


Let us now demonstrate how QS helps ensure that pressure-driven singular currents do not generally form on rational surfaces. We can obtain the ideal current from \eqref{JxB} in the form
\begin{align}
    \J=\jpl \B + \frac{\B\times \dl p}{B^2},
\end{align}
where the parallel component $\jpl$ is not yet determined. To determine $\jpl$, we use the fact that the current must be divergence-free due to \eqref{JiscurlB}. This leads to the following consistency condition 
\begin{align}
    \BD \jpl + \B \times \dl p\cdot \dl \lbr \frac{1}{B^2} \rbr=0.
    \label{eq:BDjpl}
\end{align}
Equation \eqref{eq:BDjpl} is a magnetic differential equation for $\jpl$. Single-valuedness of $\jpl$ leads to the following Newcomb constraint on every closed field line
\begin{align}
    \oint d\ell\: \B \times \dl p\cdot \dl \lbr \frac{1}{B^2} \rbr=0.
    \label{eq:Newcomb_general}
\end{align}
Without a continuous symmetry such as axisymmetry, the Newcomb condition is generally not satisfied on all rational surfaces, permitting the formation of singular currents in 3D ideal MHD equilibria. 

The QS condition \eqref{eq:2term_QS}, implies that 
\begin{align}
    \B \times \dl p\cdot \dl \lbr \frac{1}{B^2} \rbr=\BD\lbr F(\psi) \frac{p'(\psi)}{B^2} \rbr,
    \label{eq:QS_2term_1/B^2}
\end{align}
which satisfies the Newcomb constraint. This also allows us to solve for $\jpl$ up to a flux function $H'(\psi)$
\begin{align}
    \jpl = -H'(\psi) -F(\psi) \frac{p'(\psi)}{B^2}.
    \label{eq:jpl_exp}
\end{align}
We note here that the pressure-driven term in \eqref{eq:jpl_exp} constitutes the Pfirsch-Schluter current, whereas the flux function, $H'(\psi)$, encompasses all other non-pressure-driven current sources. Since a $\BD$ has been lifted to obtain $\jpl$ from \eqref{eq:BDjpl}, singular Dirac-Delta currents \citep{loizu2015_current_sheets, huang2022numerical_delta_function,huang2023_pressure_driven_HKT,rodriguez2021islands} are possible as well on rational surfaces. We plan to address the singular currents in the future.

In the case of axisymmetry, $\jpl$ can be written as an elliptic Laplacian-like operator acting on the poloidal flux, leading to a GS equation. We would now like to derive a QS-GS equation in analogy with axisymmetry. The general QS-GSE \citep{burby2020} (discussed in Appendix \ref{app:BKM}) being too complicated, we shall use the reduced model for ideal MHD derived by \cite{strauss1997reduced_near_vacuum} in the remainder of this section. Strauss' model assumes that the magnetic field is approximately given by
\begin{align}
    \B = \B_v + \dl A\times \B_v, \quad \B_v= \dl \chi,
    \label{eq:Strauss_B_form}
\end{align}
where $\B_v$ is a vacuum field, $A$ is an $O( L_\perp/L_\parallel)$ function that is analogous to the poloidal flux in the axisymmetric case. The fundamental assumption is that the gradients along $\B_v$ are smaller than those perpendicular to it. The form \eqref{eq:Strauss_B_form} ensures that $\B$ is divergence-free to first-order in $L_\perp/L_\parallel$. We shall assume that plasma beta is first order in $ L_\perp/L_\parallel$.

Calculating $\jpl$ by taking the curl of \eqref{eq:Strauss_B_form} and substituting in \eqref{eq:jpl_exp} then leads to the following elliptic PDE for $A$
\begin{align}
    -\dlts A + F(\psi) \frac{p'(\psi)}{B^2} +H'(\psi)=0, \quad \dlts A= \frac{1}{B_v^2}\dl\cdot \lbr B_v^2 \dl A \rbr.
    \label{eq:QS-GSE_Strauss}
\end{align}
Thus, \eqref{eq:jpl_exp} is equivalent to an elliptic QS-GSE, which reduces to the standard GS in the axisymmetric limit, as shown in Appendix \ref{app:BKM}. This follows immediately from  $\B_v \propto \dl \phi, B_v^2 \propto 1/R^2, A\propto -\psi$ in standard axisymmetric cylindrical coordinates. Thus, the $(F/B^2) p', H'$ terms reduce to the standard $R^2 p', I I'$ terms in the axisymmetric Grad-Shafranov equation.

\section{The Quasisymmetric High-Beta Stellarator model}
\label{sec:QS-HBS_model}
In the previous Section, we have shown that perfect QS helps to integrate out one hyperbolic characteristic of ideal MHD \citep{grad1971plasma}. Therefore, the Hamada condition \citep{helander2014} is automatically satisfied, which leads to a Pfirsch-Schluter current that is non-singular on rational flux surfaces. However, there remains another hyperbolic characteristic of ideal MHD that is related to $\BD\psi=0$. In the absence of a continuous symmetry, nested flux surfaces do not exist, and islands and chaotic fields are formed generically throughout the plasma volume. 

In this Section, we shall primarily focus on the HBS model by choosing the lowest-order vacuum field to be purely toroidal, which leads to an analytically tractable model. We shall show that the assumption of QS also helps support nested flux surfaces.

\subsection{Derivation of the QS-HBS model}\label{sec:QSHBS_derivation}

Henceforth, we assume that the aspect ratio is large and that the leading-order magnetic field is a purely toroidal vacuum field. It is then convenient to use the standard right-handed cylindrical coordinates $(R, Z, \phi)$, with unit vectors $(\bm{e_R},\bm{e_Z},\bm{e_\phi})$
\begin{align}
\bm{r}=R\bm{e_R}+ Z \bm{e_Z}, \quad \bm{e_\phi}=\bm{e_R}\times  \bm{e_Z}, \quad \dl \phi =\frac{1}{R}\bm{e_\phi}.
\label{eq:cyl_coords}
\end{align}
The inverse aspect ratio $\ep$, defined as the ratio of of the minor radius $r_0$, and the major radius $R_0$, is our expansion parameter, i.e.,
\begin{align}
    \ep\equiv \frac{r_0}{R_0} \ll 1.
    \label{eq:ep_def}
\end{align}
However, assuming a purely toroidal vacuum field to lowest order implies that the magnetic axis will remain close to a planar ring whose normal does not rotate around itself. Therefore, we can \textit{only} access QA for which $N=0$  \citep{landreman2019,rodriguez2023constructing} and
\begin{align}
    F(\psi)= \frac{G(\psi)}{\iota(\psi)} \quad \quad \text{(quasiaxisymmetry)}.
    \label{eq:F_for_QA}
\end{align}
Following \cite{freidberg2014idealMHD}, we now expand the various physical quantities in formal power series of $\ep$,
\begin{align}
    \B =\B_0+\ep \B_1 +\dots,  \;\; B =B_0 +\ep B_1 +\dots, 
    \;\; p=\ep p_1 +\dots \;\; \J = \ep \J_1+\dots,
\end{align}
assuming $B_0$ to be a constant, and the plasma beta and currents are first-order in $\ep$. All quantities such as $\B_1,B_1,p_1$ etc are assumed to be $O(1)$. It is convenient to normalize all length scales with the minor radius $r_0$. Thus,
\begin{align}
    R=\frac{1}{\ep}+x ,\quad Z=y, \quad \dl \phi = \bm{e_\phi}\; \ep(1+\ep x)^{-1},
    \label{eq:lowest_order_geom}
\end{align}
where, $(x,y)$ are order unity coordinates along $\bm{e_R}$ and $\bm{e_Z}$. Since the lowest-order magnetic field is assumed to be a toroidal vacuum field, $G\sim R_0 B_0$, and 
\begin{align}
    \B = \frac{1}{\ep}G_{-1} \dl \phi + O(\ep)= B_0 \bm{e_\phi}+ O(\ep), \quad G_{-1}=B_0.
    \label{eq:B_approx_toroidal}
\end{align}
The magnetic field is thus divergence-free and curl-free to $O(\ep)$. Moreover, $\B_0\cdot \dl= O(\ep)$ from 
\eqref{eq:lowest_order_geom} as required from $L_\perp/L_\parallel\sim \ep$. Thus, the (normalized) gradients naturally split into 
\begin{align}
    \dl =\dlr + \ep \dl_\phi,\quad  \dlr \equiv \bm{e_R}\del_x + \bm{e_Z}\del_y, \quad \dl_\phi \equiv \bm{e_\phi}(1+\ep x)^{-1} \del_\phi.
    \label{eq:different_nablas}
\end{align}
This completes the description of the lowest-order magnetic field and its associated coordinate system. Next, we analyze the ideal MHD system \eqref{eq:ideal_MHD} and the quasisymmetry condition \eqref{eq:2term_QS}, order by order. 
Since $B=B_0+O(\ep)$ and $\BD\sim \ep$, the QS condition only imposes a constraint at $O(\ep^2)$ and higher. Except for the treatment of QS, all other details of the expansion are given in \citep{freidberg2014idealMHD}, so we only include the essential results here. We note that one must be careful with the operator $\nabla_\phi$ as defined in \eqref{eq:different_nablas} since it generates $x$ dependent terms in higher orders to account for the $1/R$ term in $\dl\phi$.

Proceeding to $O(\ep)$, we find
\begin{subequations}
    \begin{align}
    \dlr \cdot \B_1=0 \label{eq:divB1},\\
    \J_1\times \B_0 = \dlr p_1, \label{eq:force_bal1}\\
    \J_1=\lbr \dl \times \B\rbr_1. \label{eq:curlB1}
\end{align}
 \label{eq:ideal_MHD_O(ep)}
\end{subequations}
It can be shown \citep{freidberg2014idealMHD} that the divergence-free condition \eqref{eq:divB1}, leads to
\begin{align}
   \frac{\B_1}{B_0} =  -\lbr x +b_{\phi 1}\rbr\bm{e_\phi} + \dl \lbr\frac{A_1}{B_0}\rbr \times \bm{e_\phi},
    \label{eq:B1_form}
\end{align}
where the $A_1$ term defines a poloidal magnetic field with $A_1$ being the stream function. The $b_{\phi 1}$ term on the right of the above equation, which originates from the pressure-induced corrections to the $1/R$ vacuum field, can be determined in terms of plasma beta using the force-balance condition \eqref{eq:force_bal1} and the definition of current $J_1$ \eqref{eq:curlB1},
\begin{align}
   b_{\phi 1}(\psi)= \frac{p_1(\psi)}{B_0^2}.
    \label{eq:bphi1}
\end{align}
Taking into account the orthogonality of the two terms in \eqref{eq:B1_form}, it follows that the field strength $B\approx B_0+\ep B_1$, where 
\begin{align}
    \frac{B_1}{B_0}=- \lbr x + b_{\phi 1}(\psi) \rbr.
    \label{eq:modB1_expr}
\end{align}
Finally, taking the curl of $\B$ to first-order, we obtain the current,
\begin{align}
    \frac{\J_1}{B_0}= -\bm{e_\phi} \dlrs \lbr\frac{A_1}{B_0}\rbr-\dlr b_{\phi 1}\times \bm{e_\phi}, \quad \dlrs \equiv \del_x^2 +\del_y^2.
    \label{eq:J1}
\end{align}
As discussed in the previous Section, the parallel component of $\J_1$ can be determined through a magnetic differential equation of the form \eqref{eq:BDjpl}. This requires us to go to second order in $\ep$.

From the second order, we find the two fundamental HBS equations,
\begin{subequations}
    \begin{align}
    \dpl \psi=0 \label{eq:HBS1_BDpsi}, \\
    \dpl \lbr \dlrs \lbr \frac{A_1}{B_0}\rbr\rbr=-2 \frac{d b_{\phi 1}}{d\psi}\del_y \psi \label{eq:HBS2_BDjpl},
\end{align}
\label{eq:HBS_system}
\end{subequations}
where we have used
\begin{align}
    \BD=\ep B_0 \dpl, \quad \dpl\equiv  \del_\phi -\left\{\lbr\frac{A_1}{B_0} \rbr, \;\;\right\}_{(x,y)}.
    \label{eq:BD_dpl_def}
\end{align}
The Poisson bracket appearing in \eqref{eq:BD_dpl_def} is defined in the usual way as
\begin{align}
    \left\{\cA_1, \;\;\right\}_{(x,y)}= \bm{e_\phi}\times \dlr \cA_1 \cdot \dlr.
    \label{eq:PB_def}
\end{align}
The system \eqref{eq:HBS_system} is much simpler than the full ideal MHD system but is still highly nonlinear. 

Now, let us turn to the QS condition \eqref{eq:2term_QS}. We can obtain $G(\psi)$ and $F(\psi)$ from their definitions \eqref{eq:Boozer_G_and_I} and \eqref{eq:F_for_QA}  as
\begin{align}
    \frac{G(\psi)}{B_0}= \oint \frac{d\phi}{2\pi} R\frac{\B}{B_0} \cdot \bm{e_\phi} \approx \frac{1}{\ep} -b_{\phi 1}, \quad  \frac{F(\psi)}{B_0}= \frac{1}{\ep\; \iota(\psi)} \lbr 1 -\ep b_{\phi 1}\rbr
    \label{eq:G_F_approx_def}.
\end{align}
The QS condition \eqref{eq:2term_QS} then implies that 
\begin{align}
    \bm{e_\phi}\times \dlr \psi_F \cdot \dlr B_1 = B_0\dpl B_1,
    \label{eq:QS2term_O1}\\
    \psi_F=\int d\psi \frac{B_0}{\ep F(\psi)}= \int d\psi\; \iota(\psi)(1+\ep b_{\phi_1}).
\end{align}
It follows from the expression \eqref{eq:G_F_approx_def} of $F(\psi)$ that to the lowest order in $\ep$, $\psi_F$ is equal to the poloidal flux 
\begin{align}
     \psi_F \approx \psi_p\equiv  \int \iota(\psi) d\psi.
     \label{eq:psi_p_def}
\end{align}
Now, substituting $B_1$ from \eqref{eq:modB1_expr} into \eqref{eq:QS2term_O1}, and using \eqref{eq:BD_dpl_def} and \eqref{eq:PB_def} we can rewrite the QS constraint as
\begin{align}
    \del_y \lbr \psi_F+A_1 \rbr =0.
    \label{eq:QS_simplified}
\end{align}
Lifting the partial $y$ derivative in \eqref{eq:QS_simplified}, we obtain
\begin{align}
    \psi_F + A_1= B_0 \mfa(x,\phi)
    \label{eq:psip_A1_ca_relation}.
\end{align}

Equations \eqref{eq:HBS_system} and \eqref{eq:psip_A1_ca_relation} constitute our quasisymmetric HBS system. To proceed further, it is convenient to introduce normalized variables $(\beta, \Psi,\cA)$ such that
\begin{align}
    \Psi=\frac{\psi_F}{B_0} ,\quad \cA=\frac{ A_1}{ B_0}  , \quad \beta=\frac{p_1}{B_0^2} .
    \label{eq:norm_var}
\end{align}
In terms of the normalized variables, together with the definitions
\begin{align}
    \dpl \equiv \del_\phi -\{\cA,\;\;\}_{(x,y)}, \quad \dlrs \equiv  \del_x^2 +\del_y^2,
    \label{eq:dpl_dlrs_def}
\end{align}
the QS-HBS model reads
\begin{subequations}
    \begin{align}
        \dpl \Psi &=0, \qquad \qquad \qquad \qquad \;\; \text{(nested flux surface condition)} \label{eq:QSHBS_1_BDpsi} \\
        \dpl \dlrs \cA &= -2 \frac{d \beta}{d\Psi}\del_y \Psi, \qquad \qquad \quad \text{(parallel current condition)}
        \label{eq:QSHBS_2_BDjpl}\\
        \Psi +\cA &=\mfa(x,\phi). 
        \qquad \qquad \qquad \text{ (relation of $\Psi$ and $A$ from QS)}
        \label{eq:QSHBS_3_psiA_reln}
    \end{align}
    \label{eq:QSHBS}
\end{subequations}
Once these equations are solved, we can calculate $\B,\J$ to $O(\ep)$ through
\begin{subequations}
    \begin{align}
    \frac{\B}{B_0} &= \bm{e_\phi} +\ep \lbr -(x+\beta)\bm{e_\phi}+\dlr \cA \times \bm{e_\phi} \rbr, \label{eq:B_form}\\
    \frac{B}{B_0} &=1-\ep \lbr x + \beta \rbr, \label{eq:modB_form}\\
    \frac{\J}{B_0} &=\ep \lbr -\bm{e_\phi} \dlrs \cA - \dlr \beta \times \bm{e_\phi}\rbr.
    \label{eq:J_form}
\end{align}
\label{eq:BBJ_forms}
\end{subequations}
This completes our derivation of the basic equations that govern near-axisymmetric quasisymmetric reduced MHD. These are nonlinear equations for $(\Psi,\cA)$, with single-valuedness of $(\Psi,\cA)$ as boundary conditions in the $(x,y,\phi)$ coordinates. The pressure term $\beta=p_1/B_0^2$ is a free input function. The function $\mathfrak{a}(x,\phi)$ is also an input function, but there are some constraints on it that we will discuss in Section \ref{sec:QSGSE_consistency} and Appendix \ref{app:consistency}.


In the next Section, we proceed with the QS-HBS system \eqref{eq:QSHBS} and obtain a QS-GSE for the function $\Psi$.

\subsection{Derivation of the quasisymmetric Grad-Shafranov equation}\label{secQS_GS_derivation}
The QS-GSE was obtained in full generality in \citep{burby2020}. However, it is hard to use and somewhat opaque due to the complexities arising from the 3D nature of the metric coefficients and the additional consistency conditions. The large aspect-ratio expansion allows us to cut through these complications and reduce the QS-GSE to a simple form, thus highlighting the essential role of geometry and QS. 

We begin by noting that the definition of $\dpl$ \eqref{eq:dpl_dlrs_def}, and the relation between $\Psi$ and $\cA$ \eqref{eq:QSHBS_3_psiA_reln} yields the identity
\begin{align}
    \dpl x =-\del_y \Psi,
    \label{eq:id_QS}
\end{align}
which is the large aspect-ratio limit of the 2-term QS formula \eqref{eq:2term_QS}. Identity \eqref{eq:id_QS} implies that the parallel current equation \eqref{eq:QSHBS_2_BDjpl} can be written as 
\begin{align}
    -\dlrs \cA + 2 x\frac{d\beta}{d\Psi} + H'(\Psi)=0.
    \label{eq:preQSGSE}
\end{align}

To obtain a single nonlinear equation for $\Psi$ we now eliminate $\cA$ from \eqref{eq:preQSGSE} using \eqref{eq:QSHBS_3_psiA_reln}. We then obtain the QS-GSE
\begin{align}
     \dlrs \Psi+ 2 x\frac{d\beta(\Psi)}{d\Psi} + H'(\Psi)-\mfa_{,xx}=0.
    \label{eq:QSGSE}
\end{align}
Here and elsewhere, we use the notation $f_{,x}$ to denote the $x$ derivative of $f$. 

Next, we substitute $\cA=\mfa -\Psi$ in the $\dpl \Psi=0$ equation to find
\begin{align}
    \Psi_{,\phi} -\mfa_{,x}\Psi_{,y}=0.
    \label{eq:uDpsi}
\end{align}
Using the method of characteristics, we can obtain the following exact solution to \eqref{eq:uDpsi} in the form of a traveling wave (TW)
\begin{align}
    \Psi=\Psi(x,\xi), \quad \xi= y + \int d\phi\; \mfa_{,x}.
    \label{eq:trav_wave_Psi}
\end{align}
To understand the physical meaning of the quantity $\xi$, we now construct the QS vector $\u$ as defined in \eqref{eq:QS_u}. As shown in Appendix \ref{app:BKM},
\begin{align}
    \u= \frac{1}{\ep\; \iota}\lbr \bm{e_\phi} (1+\ep x)-\ep \mfa_{,x}\bm{e_Z} \rbr, \quad \uD = \frac{1}{\iota} \lbr \del_\phi - \mfa_{,x}\del_y\rbr
    \label{eq:u_and_uD}.
\end{align}
From \eqref{eq:u_and_uD} we find that $\uD \xi=\uD x=0$. Thus, the symmetry vector $\u$ lies on surfaces of constant $\xi$ and $x$, thereby ensuring that $\uD \Psi(x,\xi)=0$. Therefore, $\xi$ denotes one of the Clebsch variables for $\u$, the other being $x$. Moreover, since the symmetry lines must close on themselves on a torus, $\xi$ must be periodic in $\phi$. 

We now look into the relation of the quasisymmetry vector, $\u$, with the Killing vectors of 3D Euclidean space that generate rotations and translations \citep{burby2020}. Following \citep{burby2020}, we define 
\begin{align}
    \u_{\text{HBS}}\equiv \iota \u,
    \label{eq:u_HBS}
\end{align}
which is equivalent to choosing the poloidal flux instead of toroidal flux in the expression for $\u$ \eqref{eq:QS_u}. Comparing the symmetry vectors corresponding to axisymmetry, helical symmetry, and the QA symmetry in the HBS 
\begin{align}
    \u_{\text{AS}}= R\bm{e_\phi}, \quad \u_{\text{HS}}= R\bm{e_\phi}-l\; \bm{e_Z}, \quad  \u_{\text{HBS}}\approx  R\bm{e_\phi} - \mfa_{,x} \bm{e_Z},
    \label{eq:various_symmetries}
\end{align}
we find that they have the form of a linear combination of a rotation in $\phi$ and a translation in $Z$. We note that in a straight cylinder with helical symmetry, it is customary to use the poloidal angle $\vartheta$ to denote the angle of rotation, and $\phi$ is along $Z$. We have chosen not to do so in order to express the close relationship between these vectors. 

From \eqref{eq:various_symmetries}, we observe that the translation in $Z$ is zero for axisymmetry, constant $l$ for helical symmetry, and periodic (with zero average) in HBS. As shown in Appendix \ref{app:BKM}, although the axisymmetry and helical symmetry vectors are Killing, the HBS symmetry vector is not Killing for a generic $\phi$ dependent $\del_x\mfa(x,\phi)$. Thus, the QA symmetry of the HBS model is indeed a hidden symmetry and not an isometry of the 3D Euclidean space.

\subsection{QS-GSE and consistency conditions}
\label{sec:QSGSE_consistency}

In the case of axisymmetry, $\mfa=0$, which implies $\Psi_{,\phi}=0$. The QS-GSE \eqref{eq:QSGSE} then reduces to the axisymmetric GSE as expected. However, in the nonsymmetric case with $\mfa\neq 0$, it is not obvious that the QS-GSE equation \eqref{eq:QSGSE}, and the $\dpl\Psi=0$ condition \eqref{eq:uDpsi}, are consistent. The conflict lies in the fact that the nonsymmetric terms can enter $\Psi$ only through $\mfa_{,x}$ in the form of a TW given in \eqref{eq:trav_wave_Psi}. It is not immediately obvious that a TW solution will satisfy the QS-GSE \eqref{eq:QSGSE} for a general $\mfa(x,\phi)$. 

As shown in Appendix \ref{app:consistency}, a self-consistent solution is obtained if $\mfa(x,\phi)$ is of the form
\begin{align}
    \mfa_{,xx}=0 \quad \Rightarrow \quad \mfa(x,\phi)= \bar{\mfa}(\phi)+  x\; Y'(\phi),
    \label{eq:axx_is_0}
\end{align}
where, $\bar{\mfa}(\phi),Y(\phi)$ are periodic functions of $\phi$. Note that in axisymmetry $(\del_\phi=0)$, $\mfa$ must be of the form
\begin{align}
    \mfa= a_0 + a_1 x,
\end{align}
where $a_0,a_1$ are constants. The constants can then be absorbed through a simple redefinition of the current and pressure profiles in \eqref{eq:QSGSE}. Thus, $\mathfrak{a}$ can be set to zero in the axisymmetric limit with no loss of generality.

The rather strong restriction on $\mfa(x,\phi)$, given by \eqref{eq:axx_is_0}, implies that
we need to solve the following equations for $\Psi$, subject to single-valuedness of $\Psi$ and its derivatives
\begin{subequations}
    \begin{align}
    \dlrs \Psi+ 2 x\frac{d\beta(\Psi)}{d\Psi} + H'(\Psi)=0 \label{eq:QS-GSE-AS},\\ 
     \Psi=\Psi(x,\xi), \quad \xi= y + Y(\phi). \label{eq:TW}
\end{align}
   \label{eq:QSGSE_TW}
\end{subequations}
Thus, we have a tokamak-like GSE subject to a TW deformation. The profile functions $\beta(\Psi), H(\Psi)$ are related to plasma pressure and currents. The solution strategy is simply solving the GSE equation in $(x,y)$ space and then performing the shift $y\to y+Y(\phi)=\xi$.

Alternatively, one can shift from $(x,y,\phi)$ coordinates to $(x,\xi,\phi)$ coordinates, such that
\begin{align}
    \dpl = \at{\del_\phi}{(x,\xi)} +\{\Psi,\;\;\}_{(x,\xi)}, \quad \del_y \to \del_\xi, \quad \dlrs\to \overline{\nabla}^2_\perp \equiv \del_x^2+\del_\xi^2.
    \label{eq:dpl_dlrs_x_xi_coords}
\end{align}
The details of the transformation are provided in Appendix \ref{app:coords}. Since the QS-GSE equation \eqref{eq:QS-GSE-AS} only contains $(x,y)$ derivatives and $\mfa_{,x}$ is only a function of $\phi$, the transformed QS-GSE reads
\begin{align}
    \overline{\nabla}^2_\perp \Psi+ 2 x\frac{d\beta(\Psi)}{d\Psi} + H'(\Psi)=0.
    \label{eq:QSGSE_Trans}
\end{align}

Let us now compare our system with similar systems discussed in \cite{landreman2018a}, \cite{rodriguez2023constructing}, \cite{plunk2018}, \cite{plunk2020_near_axisymmetry_MHD} and \cite{burby2020}. An important consequence of the TW nature \eqref{eq:TW} of $\Psi$ is that the quasisymmetric deformation to the axisymmetric equilibrium is a periodic displacement purely in the vertical $\bm{e_Z}$ direction. Thus, the system \eqref{eq:QSGSE_TW} does not have a rotating ellipse-like solution near the axis \citep{landreman2018a}. It can be shown self-consistently \citep{rodriguez2023constructing} that as one approaches axisymmetry, the function $\sigma(\phi)$, which controls the rotation of the ellipse, goes to zero in agreement with our results. The non-rotating aspect is also markedly different from \cite{plunk2018} and \cite{plunk2020_near_axisymmetry_MHD}, where the quasisymmetric perturbations have a harmonic $\exp{(i N \phi)}$ phase factor (with $N$ being an integer) in both $\bm{e_R}$ and $\bm{e_Z}$ directions. 

A detailed comparison of \eqref{eq:QSGSE_TW} to the general QS-GSE system of \citep{burby2020} is carried out in Appendix \ref{app:BKM}. To summarize, we find that in the large aspect-ratio limit, we recover the generalized Grad-Shafranov equations for quasisymmetry derived by  \cite{burby2020}. In general geometry, it is not enough to solve the GSE, and \cite{burby2020} had to impose several extra compatibility constraints. On the other hand, the only constraint we had to impose to ensure compatibility is $\del^2_x\mfa(x,\phi)=0$. (The extra constraints presumably appear at higher order, which is beyond the scope of the current work.)

\subsection{QS-HBS as an integrable system}
\label{sec:QSGSE_integrability}

A magnetic field that satisfies MHD equilibrium conditions and possesses nested pressure and magnetic flux surfaces is, in principle, integrable (in the sense described precisely in \cite{duignan_meiss_burby2021integrability}). Implicitly, Hamada's condition is assumed to be satisfied by any MHD equilibrium with nested surfaces \citep{helander2014}. However, the integrability of a model obtained through a formal asymptotic expansion of 3D ideal MHD equations is not always guaranteed. In particular,
Freidberg's HBS model, which consists of two nonlinear coupled PDEs for $\cA,\Psi$, does not guarantee that Hamada's condition will be satisfied. Therefore, nestedness of flux surfaces can be assumed but not self-consistently demonstrated. Here, we show that adding the QS constraint on the HBS model reinforces integrability and allows us to construct explicit action-angle coordinates. We shall now use the $(x,\xi,\phi)$ coordinates to construct action-angle coordinates.

First, we note that the definition of $\dpl$ in \eqref{eq:dpl_dlrs_x_xi_coords} implies that the pair $(x,\xi)$ satisfy
\begin{align}
    \dpl \xi = \Psi_{,x}, \quad \dpl x = -\Psi_{,\xi}.
    \label{eq:dpl_x_dpl_xi_eqn}
\end{align}
Moreover, the 2D nature of $\Psi$ is explicit in the $(x,\xi,\phi)$ coordinates since
\begin{align}
    \dpl \Psi= \at{\del_\phi}{(x,\xi)}\Psi =0.
    \label{eq:Psi_is_2D}
\end{align}
We can interpret the $\dpl$ operator as a total derivative with respect to $\phi$ along the magnetic field line. Therefore, we can cast \eqref{eq:dpl_x_dpl_xi_eqn} as Hamilton's equations of motion,
\begin{align}
    \frac{d\xi}{d\phi} = \Psi_{,x}, \quad  \frac{d x}{d\phi} = -\Psi_{,\xi},
    \label{eq:Hamilton_x_xi_eqn}
\end{align}
where $\xi$ is the coordinate, $x$ is the conjugate momentum, and $\Psi(x,\xi)$ is the Hamiltonian. The condition \eqref{eq:Psi_is_2D} implies that the Hamiltonian is independent of time. 

Next, we perform a canonical transform from $(x,\xi,\phi)$ to $(\cI,\vartheta,\phi)$ such that
\begin{align}
    \frac{d\vartheta}{d\phi}= \Psi_{,\cI}, \quad  \frac{d\cI}{d\phi}= - \Psi_{\vartheta}.
\end{align}
Requiring that $\Psi$ be independent of $\vartheta$, we arrive at the action-angle coordinates, where
\begin{align}
    \cI = \frac{1}{2\pi}\oint_\Psi x d\xi, \quad  \frac{d\vartheta}{d\phi}= \frac{d\Psi}{d\cI}= \iota
    \label{eq:action_and_iota}.
\end{align}
It then follows that to this order of accuracy, the action $\cI$ is nothing but the toroidal flux $\psi$, $\vartheta$ is the canonical straight field-line poloidal angle, and $\iota$ is the rotational transform 
The following more explicit form of $\iota$ can be derived from \eqref{eq:action_and_iota} using $\Psi_{,x}x_{,\Psi}=1$:
\begin{align}
    \iota^{-1}= \frac{1}{2\pi}\oint_\Psi \frac{d\xi}{\Psi_{,x}}.
    \label{eq:iota_form}
\end{align}

In summary, we have shown that nested flux surfaces from the axisymmetric configuration are preserved since they are merely subjected to a nonsymmetric but periodic deformation implicitly through $\xi$. Moreover, the magnetic field-line dynamics of the QS-HBS system is that of an integrable Hamiltonian system. The action-angle coordinates of the QS-HBS corresponds to the usual straight-field line coordinates. In other words, QS not only enables us to avoid singular currents near rational surfaces (as discussed in Section \ref{sec:QS_iMHD_basics}) but also preserves the nested flux surface structure.

\subsection{Clebsch variables for the QS-HBS system}

In the last Section, we have discussed the straight-field line coordinates, which are the action-angle coordinates for the QS-HBS system. Often, in the local analysis of plasma equilibrium, such as MHD and kinetic stability, it is useful to use Clebsch variables. In this Section, we derive the expressions for the Clebsch variables, $(\Psi,\alpha,\ell)$ where $\alpha$ is the field-line label and $\ell$ is the arclength along $\B$. 

The expression for the field-line label $\alpha$ can be derived from the condition $\BD \alpha=0$, which implies
\begin{align}
    \dpl \alpha = \alpha_{,\phi} + \{\Psi,\alpha\}_{(x,\xi)}=0
    \label{eq:dpl_alpha_xi}.
\end{align}
To simplify \eqref{eq:dpl_alpha_xi}, we further change coordinates to $(\Psi,\xi,\phi)$. We have provided the necessary details of the transformation to the straight field-line coordinates in Appendix \ref{app:coords}. In $(\Psi,\xi,\phi)$ coordinates, \eqref{eq:dpl_alpha_xi} takes the simplified form
\begin{align}
    \lbr \del_\phi + \Psi_{,x}\del_\xi\rbr \alpha =0,
    \label{eq:dpl_alpha_psi}
\end{align}
whose solution can be immediately obtained as
\begin{align}
    \alpha = \phi - \int_\Psi\frac{ d\xi}{\Psi_{,x}} +O(\ep).
    \label{eq:alpha_form}
\end{align}
Here, we have used the fact that $\Psi$ is independent of $\phi$ at a fixed $\xi$. We can easily check that $\alpha = \phi - \iota^{-1}\vartheta$ from the Hamilton's equations for the field-lines,
\begin{align}
    \frac{d\xi}{\Psi_{,x}}=\frac{dx}{-\Psi_{,\xi}}= d\phi = \iota^{-1}d\vartheta.
    \label{eq:Hamilton_field_lines}
\end{align}
However, as shown in Appendix \ref{app:coords}, $\B$ as given in \eqref{eq:B_form} is equivalent to
\begin{align}
    \B/B_0= \dl \alpha \times \dl \Psi,
    \label{eq:B_Clebsch_form}
\end{align}
only if we keep the $O(\ep)$ correction to \eqref{eq:alpha_form}. The correction is needed since the $B_1/B_0$ terms do not contribute to the $\dpl$ operator but to $\B_1/B_0$. Thus, we choose the following form of $\alpha$:
\begin{align}
    \alpha = \phi - \int_\Psi d\xi\frac{B_\phi/B_0}{\Psi_{,x}} = \phi - \int_\Psi d\xi \frac{1-\ep(x+\beta)}{\Psi_{,x}}.
    \label{eq:alpha_accurate_form}
\end{align}
We note that there is a sign difference between \eqref{eq:B_Clebsch_form} and what is typically used in the stellarator literature \citep{haeseleer_flux_coordinates,helander2014} owing to the definition of $\alpha$ in \eqref{eq:alpha_form}. However, the benefit of this form of $\alpha$ and $\B$ is that it smoothly goes over to the tokamak expressions when the deviation from axisymmetry is negligible. To that end, we can recast $\alpha$ as
\begin{align}
    \alpha = \phi - q(\Psi)\int_\Psi d\vartheta\; b_\phi(\Psi,\vartheta), \quad q(\Psi)\equiv \iota^{-1}, \quad b_\phi = B_\phi/B_0.
    \label{eq:alpha_form_close2_A}
\end{align}
It is important to clarify that $(B_\phi/B_0-1)$ is only a function of $x(\Psi,\xi)$ and $\beta(\Psi)$, and independent of $\phi$. Through the canonical transform \eqref{eq:action_and_iota}, we then obtain $b_\phi=b_\phi(\Psi,\vartheta)$. Furthermore, since $b_\phi=1+O(\ep)$, $\alpha =\phi -q(\Psi)\vartheta+O(\ep)$. Thus, $\vartheta$ is indeed the straight field line poloidal angle as discussed in Section \ref{sec:QSGSE_integrability}.

Similarly, the expression for $\ell$ can be derived from the condition $\BD\ell=B$, or equivalently in $(\Psi,\xi,\phi)$ coordinates
\begin{align}
    \dpl \ell =\lbr \del_\phi +\Psi_{,x}\del_\xi \rbr \ell =\frac{1}{\ep},
    \label{eq:dpl_ell}
\end{align}
which implies that
\begin{align}
    \ell = \frac{1}{\ep}\lbr \phi + L(\Psi)+O(\ep)\rbr.
    \label{eq:ell_form}
\end{align}
The $\ep^{-1}$ factor in the expression for $\ell$ follows from the fact that we have normalized with respect to the minor radius, whereas the arclength scales with the major radius. The form of $\ell$ \eqref{eq:ell_form} also follows from the fact that in $(\Psi,\alpha,\phi)$ coordinates $\dpl =\del_\phi $. The function $L(\Psi)$ is a homogeneous solution of the operator $\dpl$.

\section{Analytic solutions of the QS-GSE: Extended Soloviev profiles}\label{sec:Extended_soloviev}

We now have in our possession the QS-HBS model, which ensures both quasisymmetry and force-balance to first-order in the LAE. As discussed in the previous Section, we can start with any LAE tokamak equilibrium and deform it to obtain a quasiaxisymmetric stellarator. In this Section, we show a simple analytic exact solution to the QS-HBS model that goes beyond the scope of the NAE. Using the VMEC code, we can verify that this solution approximates 3D MHD equilibrium with good volumetric quasiaxisymmetry. In particular, we demonstrate that as the aspect ratio becomes large, the QS error and the differences between the numerical 3D equilibrium from VMEC and the asymptotic analytical model tend to zero. 

We shall now look for a class of LAE MHD equilibrium with the following profile functions $(\beta(\Psi),H(\Psi))$,
\begin{align}
    \beta= \beta_0 + \beta_1 \Psi, \quad H= H_0 + H_1 \Psi + \frac{H_2}{2} \Psi^2,
    \label{eq:ext_Soloviev}
\end{align}
such that the QS-GSE takes the following form of a linear equation in $\Psi$,
\begin{align}
     \overline{\nabla}^2_\perp \Psi + 2x\beta_1 +\lbr H_1 + H_2 \Psi\rbr =0.
    \label{eq:QSGSE_ext_Soloviev}
\end{align}
Here, $\beta_i,H_j; i=0,1; j=0,1,2$ are constants that we can freely choose. If the quadratic term $H_2$ is identically zero, the profiles reduce to the so-called Soloviev profiles. It is straightforward to find a solution of equation \eqref{eq:QSGSE_ext_Soloviev} for a circular cross-section device:
\begin{align}
    \Psi= \frac{H_1}{H_2 J_0(\sqrt{H_2})}J_0(\sqrt{H_2}r) - \frac{H_1}{H_2} + \frac{2\beta_1}{H_2 J_1(\sqrt{H_2})}J_1(\sqrt{H_2}r)\cos\theta - \frac{2\beta_1 r}{H_2}\cos\theta.
\end{align}
Here $J_n$ are order-$n$ Bessel functions of the first kind. We combine this solution with a deformation $Y(\phi) = -0.5\sin{2\phi}$, where each poloidal plane is rigidly displaced in the vertical direction by $-Y(\phi)$.

We then use the above $\Psi$ to find the rotational transform numerically by taking the derivative of poloidal flux with respect to toroidal flux and passing the profile to VMEC. This was repeated for four different aspect ratios, namely 5, 20/3, 10, and 20, and $H_1$ was scaled by the inverse aspect ratio as $H_1 = 0.04/\epsilon$. The values of the other parameters were held constant at $H_2 = 4.8$ and $\beta_1 = 16\pi\cdot 10^{-2}$. Note that $\beta(\Psi)$ here is normalized to be $O(1)$; the true ratio of hydrodynamic to magnetic pressure is $\epsilon\beta(\Psi)$. To convert normalized quantities to SI units, we used the normalization factors $B_0 = 1~\mathrm{T}$ and $r_0 = 1~\mathrm{m}$. This choice of parameters corresponds to holding the parameters $r_0$, $B_0$, $p_1$, $H_{1,\mathrm{SI}}$ and $H_{2,\mathrm{SI}}$ constant when the QS-GSE is written in the SI system:
\begin{equation*}
    \nabla_\perp^2\Psi_\mathrm{SI} + \frac{2\mu_0 p_1}{R_0 B_0^2}x + \frac{H_{1,\mathrm{SI}}}{B_0} + \frac{H_{2,\mathrm{SI}}}{B_0}\Psi_\mathrm{SI} = 0,
\end{equation*}
where $p_\mathrm{SI}(\Psi_\mathrm{SI}) = p_0 + p_1\Psi_\mathrm{SI}$ and $H_\mathrm{SI}'(\Psi_\mathrm{SI}) = H_{1,\mathrm{SI}} + H_{2,\mathrm{SI}}\Psi_\mathrm{SI} = H(\Psi)$.

\begin{figure}
    \centering
    \includegraphics[width=0.45\textwidth]{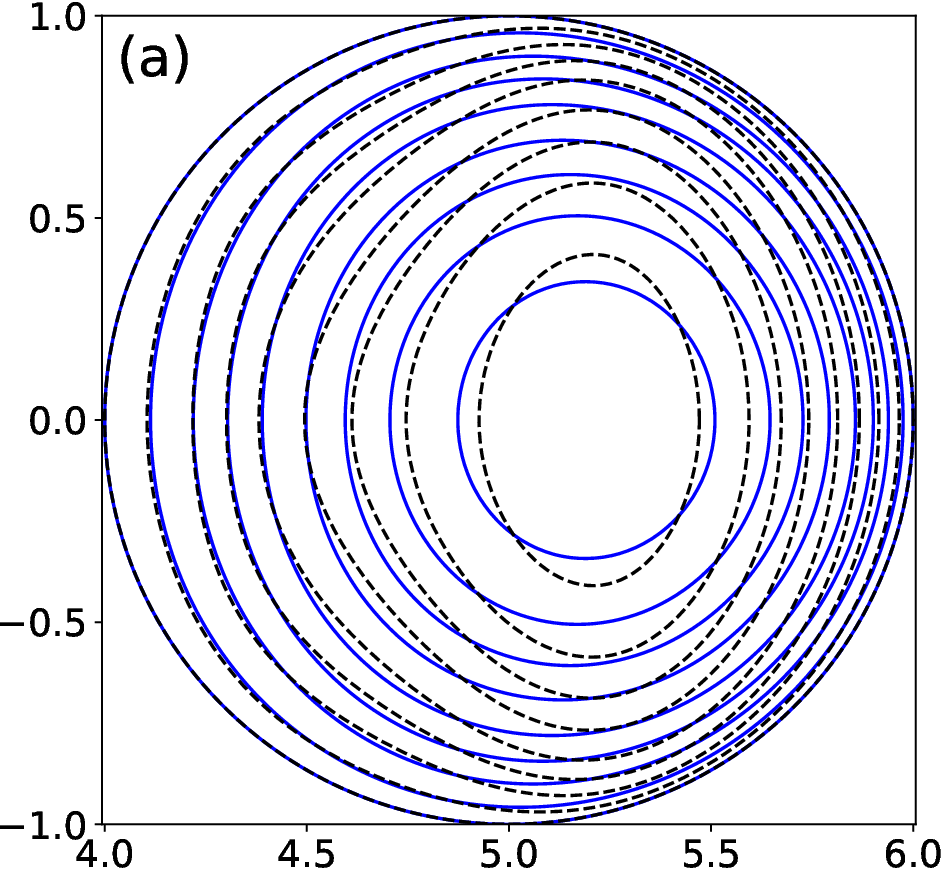}
    \includegraphics[width=0.45\textwidth]{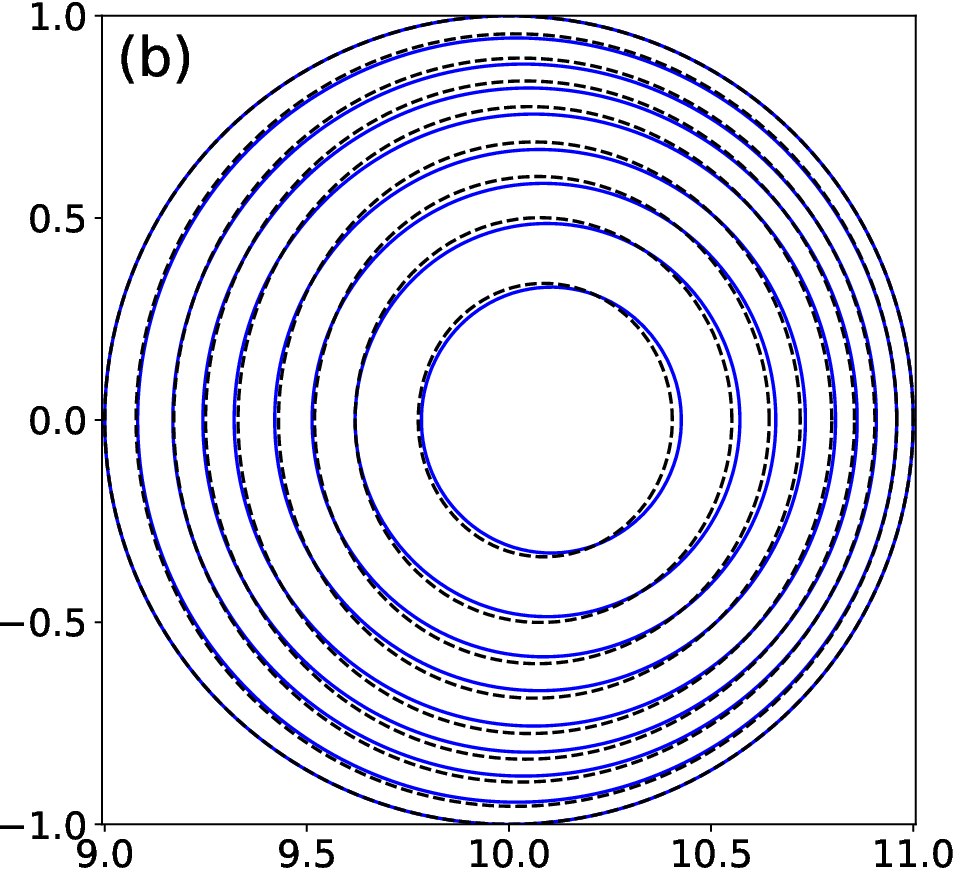}
    \caption{Approximate analytical flux surfaces (solid blue) vs. numerical flux surfaces from VMEC (dashed black) at the $\phi = 0$ poloidal plane for aspect ratios 5 (a) and 10 (b).}
    \label{fig:circ_flux_surface}
\end{figure}

\begin{figure}
    \centering
    \includegraphics[scale=0.5]{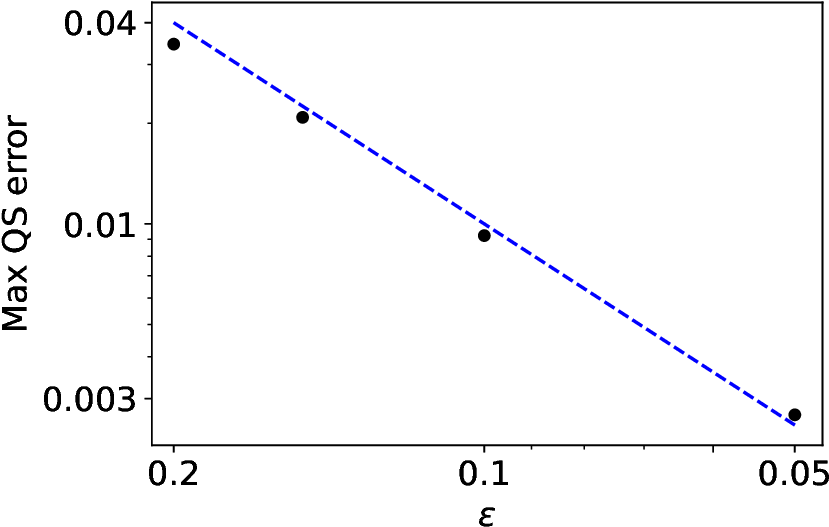}
    \caption{The maximum quasisymmetry error (black dots) scales as $\ep^2$ (dashed blue line).}
    \label{fig:circ_QSe}
\end{figure}

The flux surfaces obtained from the analytical expression and VMEC are compared in Fig. \ref{fig:circ_flux_surface}. Note that the main difference between the VMEC flux surfaces and the analytical ones is that VMEC shows a rotating ellipse effect, which cannot be captured by the analytical model, as mentioned before. Finally, the maximum QS error in the equilibrium is defined as
\begin{equation}
    \max_\psi \sqrt{\sum_{n\neq 0}\widehat{B}_{n,m}(\psi)^2\Big/\sum_{n,m}\widehat{B}_{n,m}(\psi)^2},
\end{equation}
is plotted in Fig. \ref{fig:circ_QSe} as a function of inverse aspect ratio $\ep$. Here, $\widehat{B}_{n,m}(\psi)$ is a Fourier mode of $B = |\B|$ on flux surface $\psi$. As can be seen, it scales as $\ep^2$, which is expected since the QS-HBS model is derived at order $\ep$.

\begin{figure}
    \centering
    \includegraphics[scale=0.25]{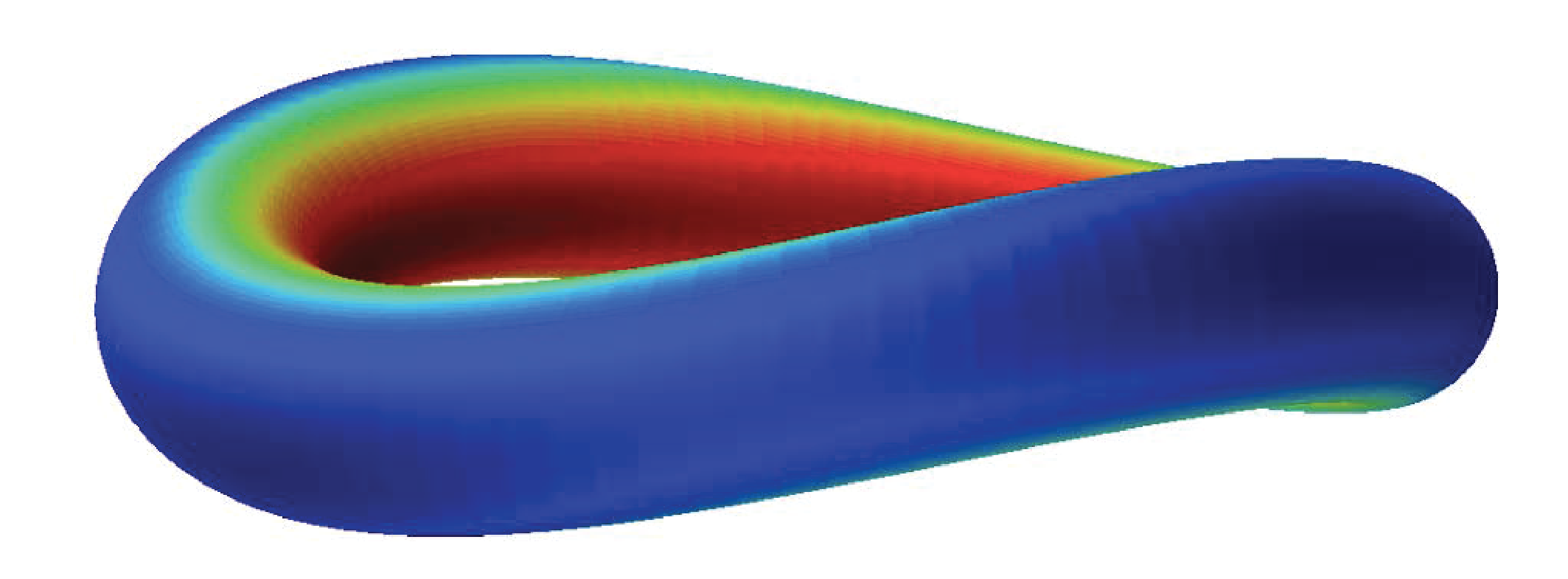}
    \caption{The deformed equilibrium with a circular cross-section and aspect ratio 5.}
    \label{fig:equil_3D_plot_circ}
\end{figure}

\section{Numerical solutions: More complex geometry}

Since equation \eqref{eq:QS-GSE-AS} matches the large aspect-ratio limit of the standard axisymmetric GSE, one can numerically solve the QS-HBS model by taking any numerical tokamak equilibrium with a large aspect ratio and applying a periodic vertical deformation as given by equation \eqref{eq:TW}.

\begin{figure}
    \centering
    \includegraphics[scale=0.25]{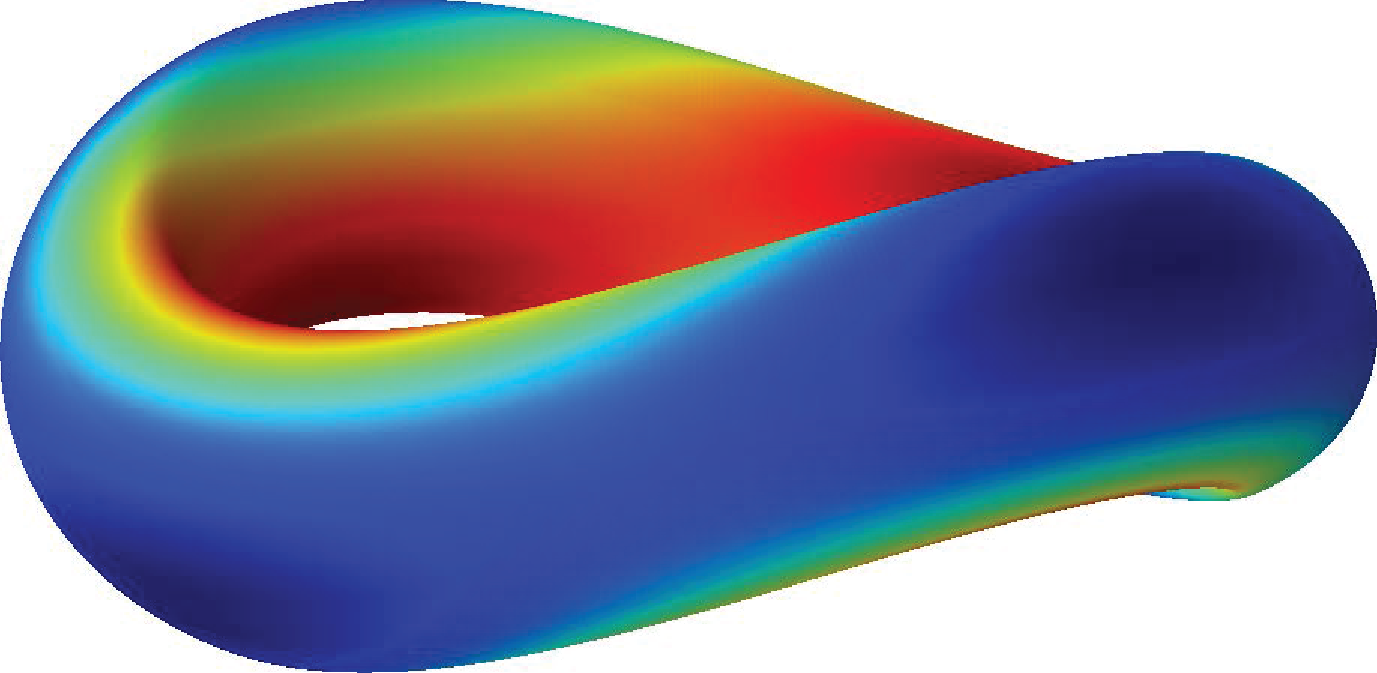}
    \caption{The deformed equilibrium with an ITER-like cross-section and aspect ratio 4.84.}
    \label{fig:equil_3D_plot}
\end{figure}

\begin{figure}
    \centering
    \includegraphics[scale=0.5]{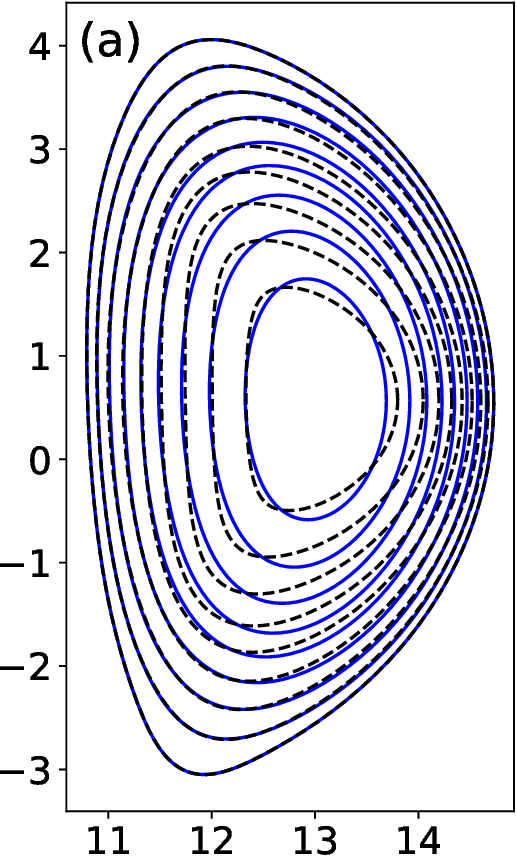}
    \includegraphics[scale=0.5]{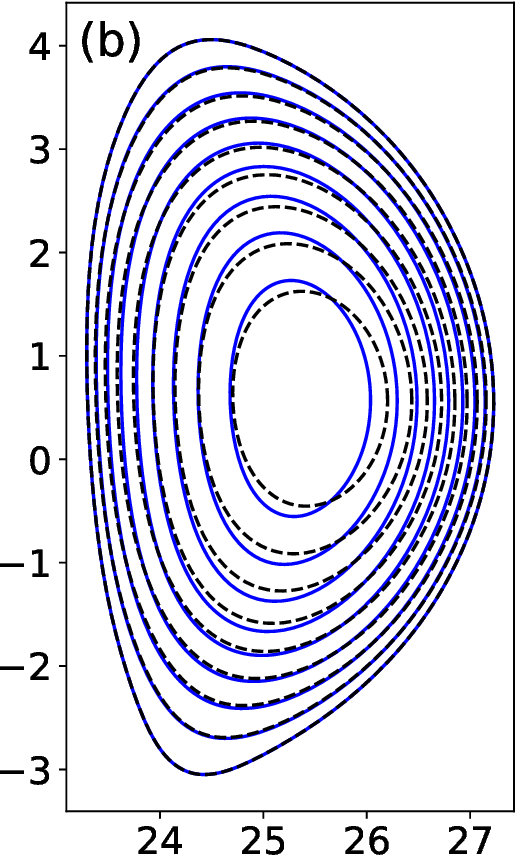}
    \caption{Flux surfaces from the axisymmetric equilibrium (solid blue) vs from the deformed equilibrium (dashed black) at the $\phi = 0$ poloidal plane for aspect ratios 5 (a) and 10 (b).}
    \label{fig:flux_surface}
\end{figure}

\begin{figure}
    \centering
    \includegraphics[scale=0.5]{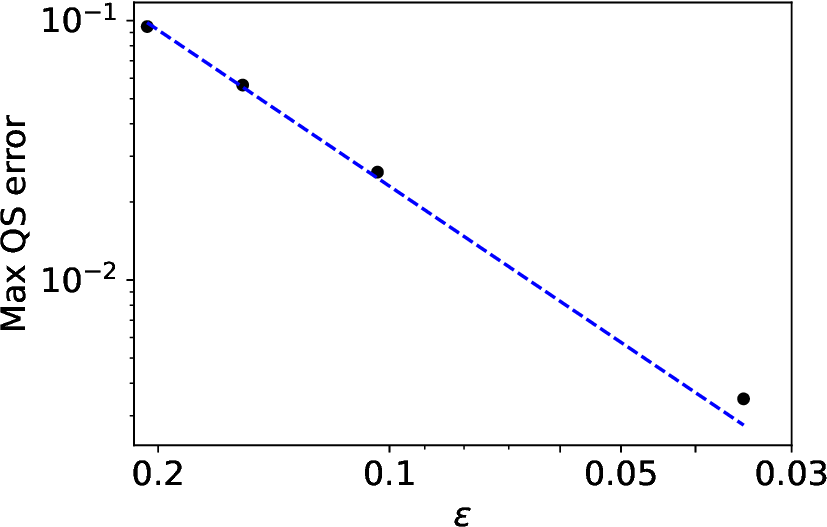}
    \caption{The maximum quasisymmetry error (black dots) again scales as $\ep^2$ (dashed blue line is $2.3\ep^2$).}
    \label{fig:QSe}
\end{figure}

With this consideration, we use VMEC to calculate an axisymmetric equilibrium with an ITER-like cross-section at aspect ratios 4.84, 6.44, 9.65, and 28.87. The toroidal current was preserved across all aspect ratios, with the $\iota$ profile varying significantly. The aspect ratios were chosen to be close to those in the previous Section except for the last one, which had to be significantly larger as the $\iota$ profile would cross 2 at aspect ratios around 20 with the chosen current profile. The presence of a low-order rational surface leads to large numerical errors in VMEC, which results in a large QS error in the deformed equilibria. In the next step, a deformation $Y(\phi) = -0.5b\sin{2\phi}$ is applied, where $b$ is the distance from the axis to the upper tip of the outermost flux surface, and the equilibria are recomputed with VMEC. The lowest aspect ratio deformed equilibrium is shown in Fig. \ref{fig:equil_3D_plot}. Figure \ref{fig:flux_surface} shows the flux surfaces from the axisymmetric and deformed equilibria plotted on top of each other. As before, the deformed equilibria show a rotating ellipse effect, which cannot be accounted for by simply displacing the axisymmetric poloidal planes vertically by $-Y(\phi)$. The maximum QS error is plotted in Fig. \ref{fig:QSe} and again scales as $\ep^2$, but is larger by a factor of $\sim 2.3$ than in the circular cross-section case.

\section{Ballooning and interchange stability}\label{sec:stability}
In the previous Sections, we developed a model for a large aspect ratio stellarator with approximate volumetric quasi-axisymmetry close to axisymmetry. We have also demonstrated good agreement of our model with actual 3D equilibria computed using the VMEC code for an aspect ratio of five and higher. These equilibria can handle plasma $\beta$ of $O(\ep)$, consistent with the high-$\beta$ stellarator model. Since high-$\beta$ equilibria are particularly sensitive to ideal interchange and ballooning instabilities, an MHD stability analysis is crucial. We note here that our stability analysis relies on standard tools like VMEC. However, it is well-known that VMEC can not accurately resolve current singularities. How such singularities can affect ideal MHD stability in the present model is a question we defer to the future.

In this Section, we perform a stability analysis of the QS-HBS equilibria obtained here with respect to interchange and ideal ballooning modes. As is well known, interchange and ideal ballooning modes arise due to the existence of regions of unfavorable curvature, i.e., $(\boldsymbol{B}\cdot \boldsymbol{\nabla}\boldsymbol{B})\cdot\boldsymbol{\nabla}p > 0$, which tends to destabilize the plasma. 

First, we calculate the Mercier (or interchange) stability quantified by the quantity $D_\mathrm{Merc}$ from VMEC. The Mercier stability criterion is
\begin{equation}
    D_\mathrm{Merc}  = -\frac{p^{'}}{\iota^{'}}V^{\dagger \dagger} - \frac{1}{4} > 0,
\end{equation}
where the quantitiy $V^{\dagger \dagger}$ is related to the magnetic well~\citep{greene1997brief}. The expression for $V^{\dagger \dagger}$ contains the contributions from different terms that arise from the magnetic well, geodesic curvature, plasma current, and magnetic shear. A positive $D_\mathrm{Merc}$ corresponds to stability against interchange modes, whereas a negative $D_\mathrm{Merc}$ implies instability.

\begin{figure}
    \centering
    \includegraphics[height= 0.34\linewidth]{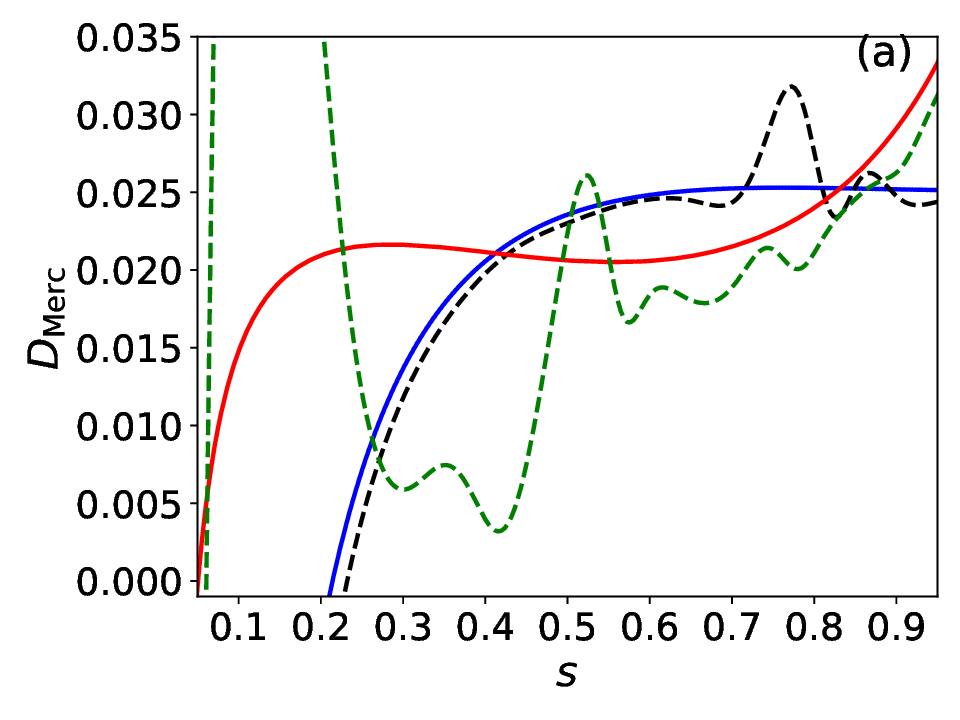}
    \includegraphics[height= 0.335\linewidth]{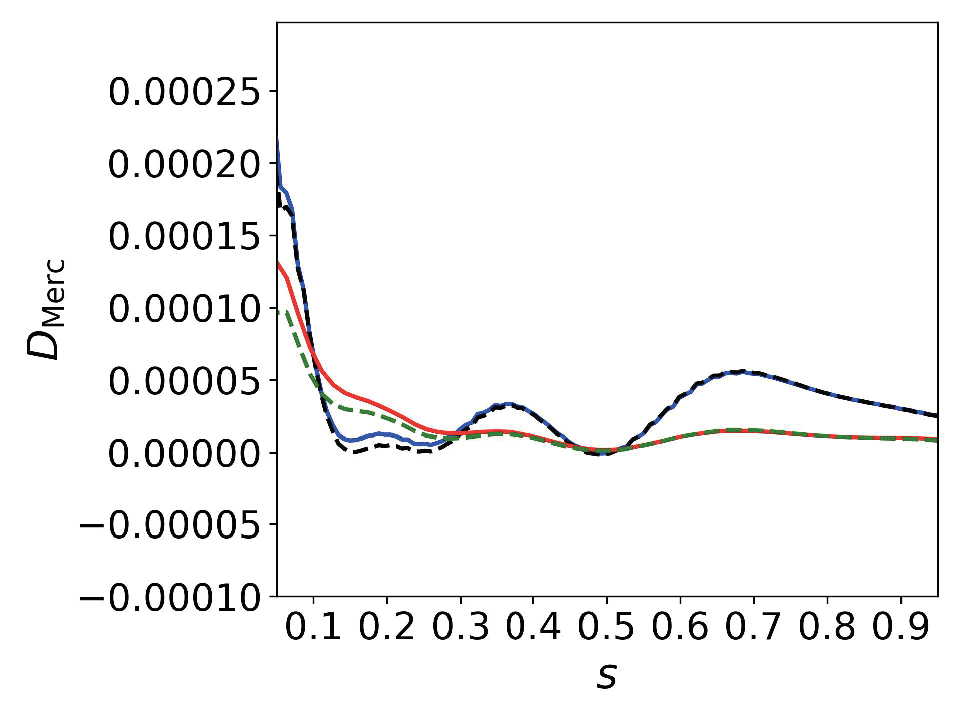}
    \caption{Mercier stability plots for the circular equilibria (a) and the ITER-like equilibria (b). The dashed curves are for the perturbed equilibria, whereas solid curves correspond to unperturbed tokamak equilibria. The red and green curves correspond to equilibria with an aspect ratio of five, and the blue and black curves correspond to an aspect ratio of ten.}
    \label{fig:DMerc_ITER_test}
\end{figure}

The radial variation of $D_\mathrm{Merc}$ for the circular and ITER-like equilibria is plotted in Figure~\ref{fig:DMerc_ITER_test}. 
Interchange modes arise in regions of low magnetic shear. For these equilibria, the magnetic shear is the lowest near the magnetic axis, which causes the circular equilibria to become Mercier unstable. Also, we note that for the ITER-like equilibria, the Mercier stability plots do not change significantly after we impose the toroidal-symmetry-breaking perturbation. 

As the equilibria analyzed are stable against interchange modes for most of the volume, we then analyze their stability against the infinite-$n$, ideal ballooning mode. The ideal ballooning mode is another curvature-driven instability that can appear in equilibria with finite magnetic shear. To calculate the stability, we numerically solve~\citep{gaur_buller_ruth_landreman_abel_dorland_2023} the ballooning eigenmode equation~\citep{connor1978shear, dewar1983ballooning}
\begin{equation}
\frac{d}{d\theta} g \frac{dX}{d \theta} + c X = \lambda f X,
\label{eqn:ideal-ballooning-equation}
\end{equation}
where $\lambda$ is the ballooning eigenvalue, and $g,c,f$ are the following normalized geometric coefficients
\begin{equation}
\begin{gathered}
    g = (\boldsymbol{b}\cdot \boldsymbol{\nabla}\theta) \frac{\lvert \boldsymbol{\nabla}\alpha\rvert^2}{B}\\
    c = \frac{1}{B^2}\frac{d(\mu_0 p)}{d\psi} \frac{2}{(\boldsymbol{b}\cdot \boldsymbol{\nabla}\theta)} (\boldsymbol{b}\times (\boldsymbol{b}\cdot \boldsymbol{\nabla}\boldsymbol{b})) \times \boldsymbol{\nabla}\alpha \\
    f = \frac{1}{(\boldsymbol{b}\cdot \boldsymbol{\nabla}\theta)} \frac{\lvert \boldsymbol{\nabla}\alpha\rvert^2}{B^3},
\end{gathered}
\end{equation}
where $\alpha = \phi - 1/\iota (\theta - \theta_0)$ is the field line label, $\theta$ is the PEST straight-field-line angle, $\phi$ is the cylindrical toroidal angle, and $\theta_0$ is the ballooning parameter. 
All lengths in the ballooning equation are normalized using $a_{\mathrm{N}}$, the effective minor radius (named $\textrm{Aminor\_p})$ in the VMEC output, and magnetic field and plasma pressure are normalized using $B_{\mathrm{N}} = 2 \psi/a_{\mathrm{N}}^2$, where $\psi$ is the toroidal flux enclosed by the boundary (without the factor of $2 \pi$).
We solve~\eqref{eqn:ideal-ballooning-equation} for the eigenvalue $\lambda = \gamma^2 a_{\mathrm{N}}^2/B_{\mathrm{N}}^2$, where $\gamma a_{\mathrm{N}}/B_{\mathrm{N}}$ is the normalized growth rate, and the ballooning eigenfunction $X$. If $\lambda > 0$, an equilibrium is unstable against the ballooning mode; otherwise, it is stable.

\begin{figure}
    \centering
    \includegraphics[height= 0.32\linewidth]{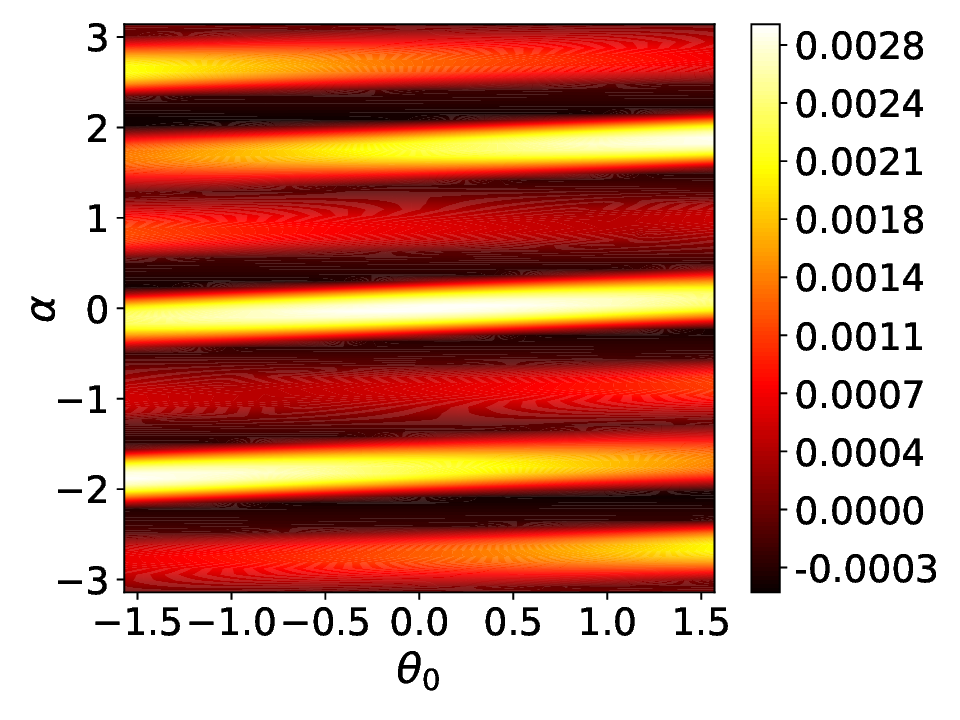}
    \includegraphics[height= 0.32\linewidth]{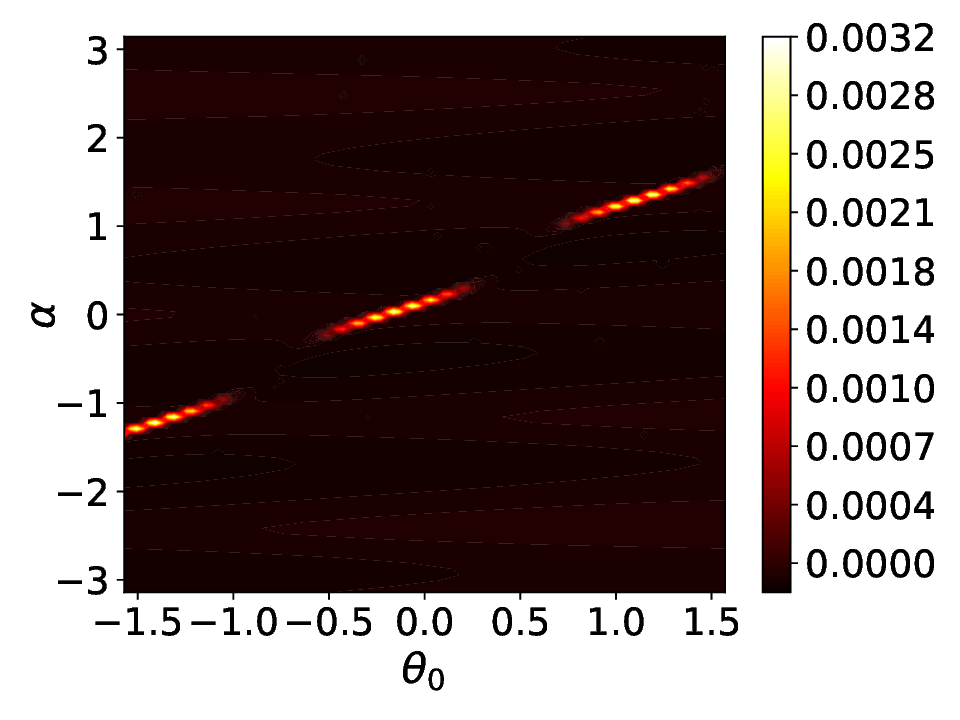}
    \caption{Ballooning eigenvalue ($\lambda$) contour plots at $\rho = 0.2$ and $\rho = 0.93$ for the perturbed aspect ratio five ITER equilibrium. The difference in the size of the ballooning unstable regions can be attributed to a large local magnetic shear in the outer region of the stellarator case. Hence, we use $n_{\alpha} = 96, n_{\theta_0} = 96$ points to accurately calculate the maximum growth rate. This process is repeated for all flux surfaces.} 
    \label{fig:alpha-theta0-contour}
\end{figure}

For tokamak geometry, we solve~\eqref{eqn:ideal-ballooning-equation} on multiple flux surfaces. For each flux surface, we scan $n_{\theta_{0}} = 96$ uniformly spaced values of $\theta_0 \in [-\pi/2, \pi/2]$ and $n_{\alpha} = 1$ on the field line $\alpha = 0$. For perturbed tokamaks, we repeat the same process on $n_{\alpha} = 96$ field lines with uniformly spaced values of $\alpha \in [-\pi, \pi]$. A typical plot of the growth rate for the perturbed ITER-like case is shown in Figure~\ref{fig:alpha-theta0-contour}.

\begin{figure}
    \centering
    \includegraphics[height= 0.32\linewidth]{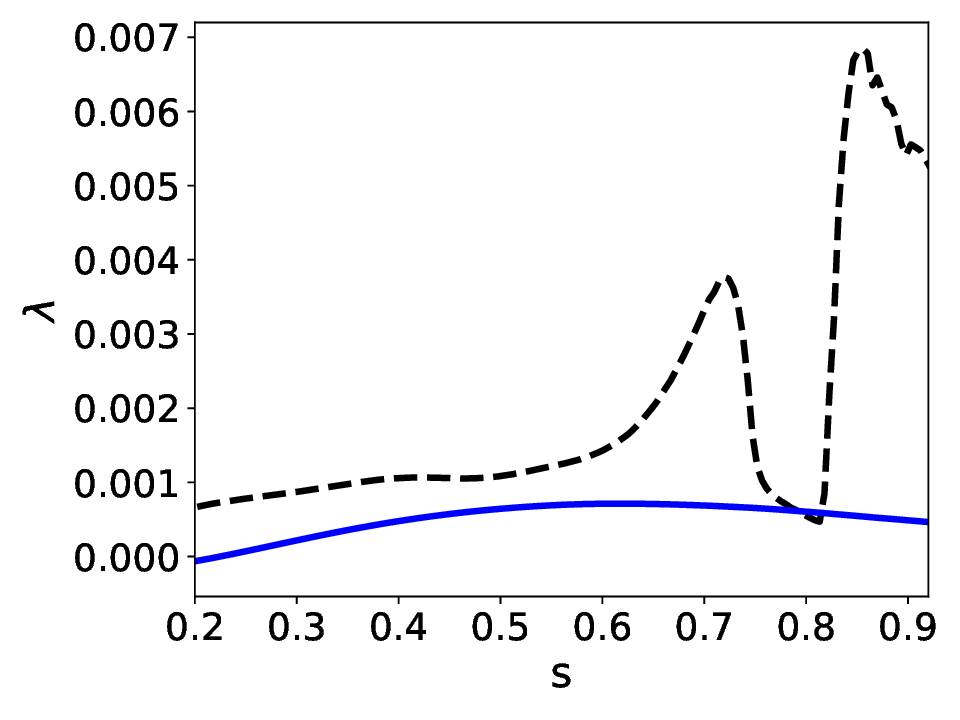}
    \caption{Ideal ballooning stability plots for the circular equilibria with an aspect ratio of ten. The dashed curves are for the perturbed equilibria, whereas solid curves correspond to unperturbed tokamak equilibria. The pressure gradient for the perturbed equilibrium is finite at the edge, which results in a finite edge growth rate.}
    \label{fig:circ_test_gamma_ball}
\end{figure}

\begin{figure}
    \centering
    \includegraphics[height= 0.32\linewidth]{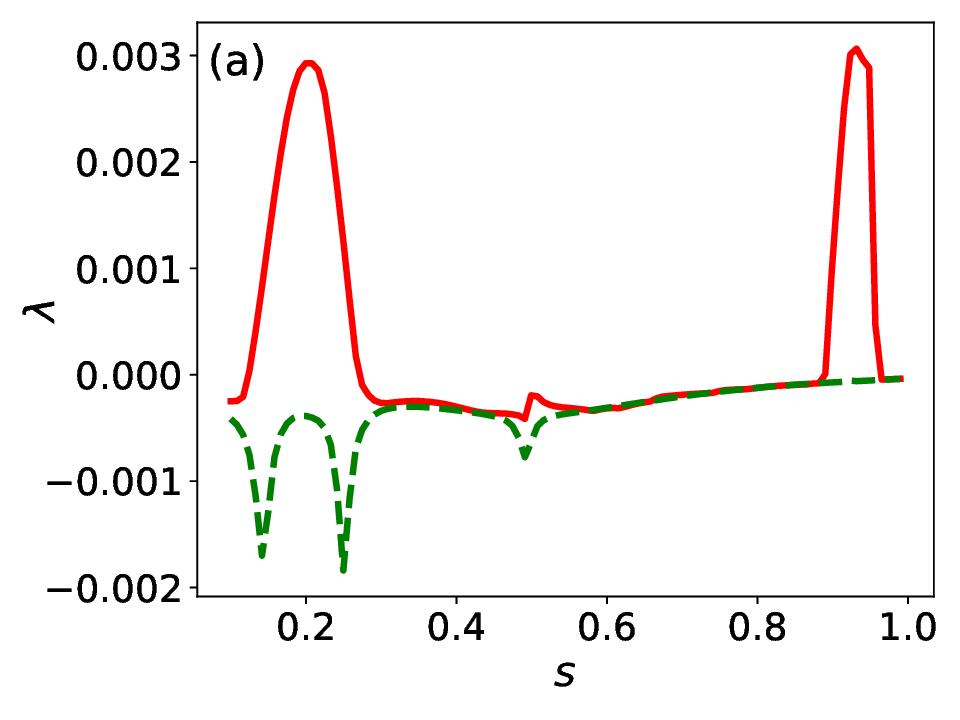}
    \includegraphics[height= 0.32\linewidth]{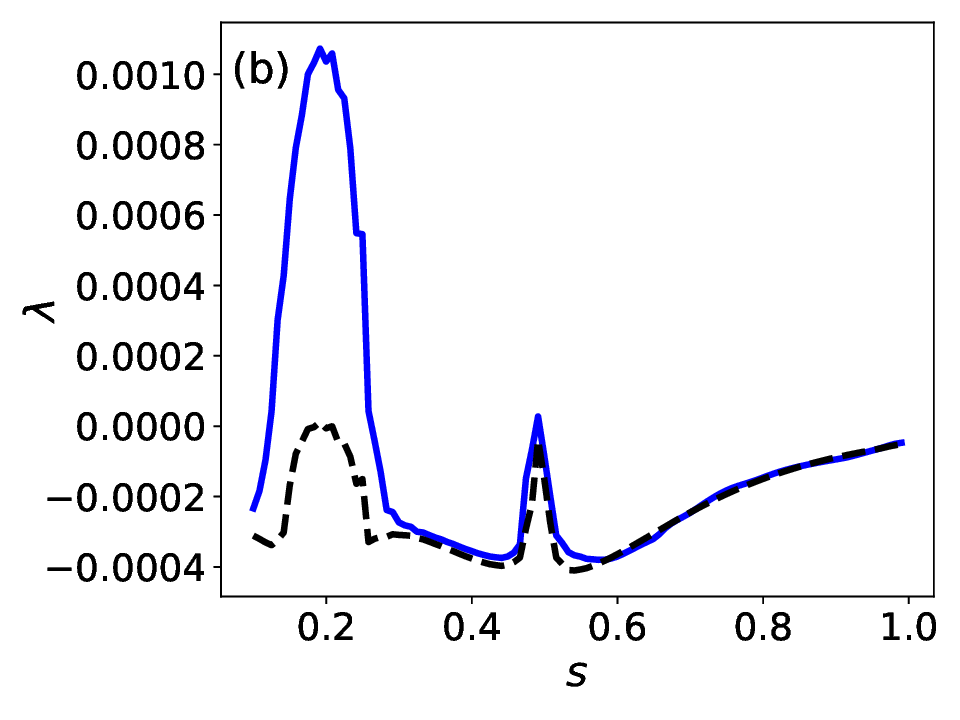}
    \caption{Ideal ballooning stability plots for the ITER-like equilibria with an aspect ratio of five (a) and ten (b). The solid curves correspond to the perturbed equilibria, whereas dashed curves correspond to unperturbed tokamak equilibria.}
    \label{fig:arbit_gamma_ball}
\end{figure}

In Figure~\ref{fig:circ_test_gamma_ball}, we present the ideal ballooning stability eigenvalue $\lambda$ as a function of the radial coordinate $s$.
For the large aspect ratio tokamak, we observe the emergence of ballooning instability after perturbing it into a stellarator. This is not surprising since a lack of shaping will, in general, destabilize the ballooning mode. We repeat this process for the perturbed ITER-like equilibrium and plot the ballooning eigenvalue in Figure~\ref{fig:arbit_gamma_ball}.
The tokamak equilibria are ballooning stable, whereas the deformed tokamak becomes ballooning unstable near the axis for both aspect ratios and near the edge for the aspect ratio five. However, because of the strong shaping, the ITER-like equilibria have significantly better ballooning stability properties after being perturbed compared to the circular tokamak case. Thus, strongly shaped QS-HBS equilibria may retain favorable stability even after the toroidal symmetry is broken. 

\begin{figure}
    \centering
    \includegraphics[height= 0.30\linewidth]{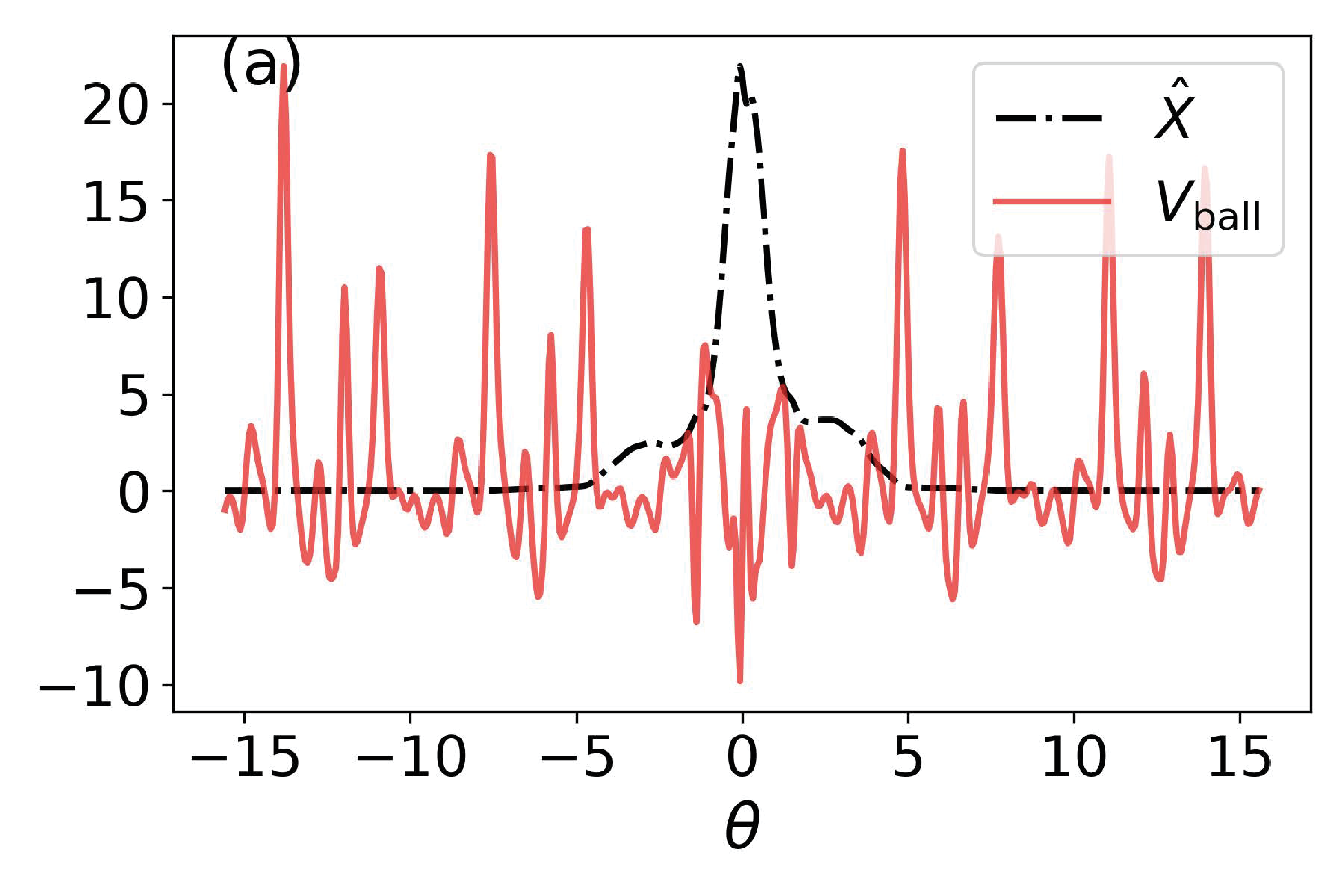}
    \includegraphics[height= 0.30\linewidth]{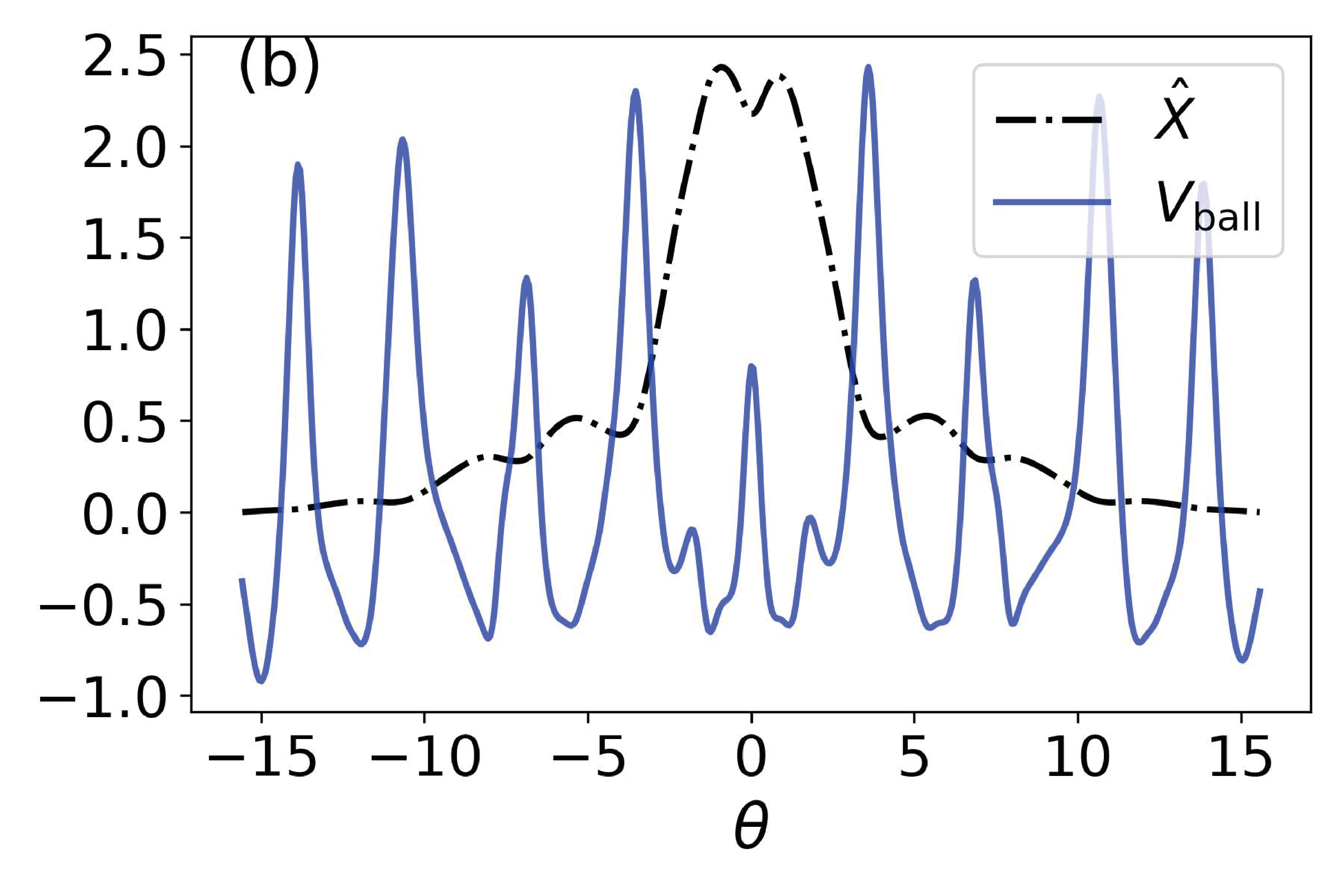} \\
    \includegraphics[height= 0.30\linewidth]{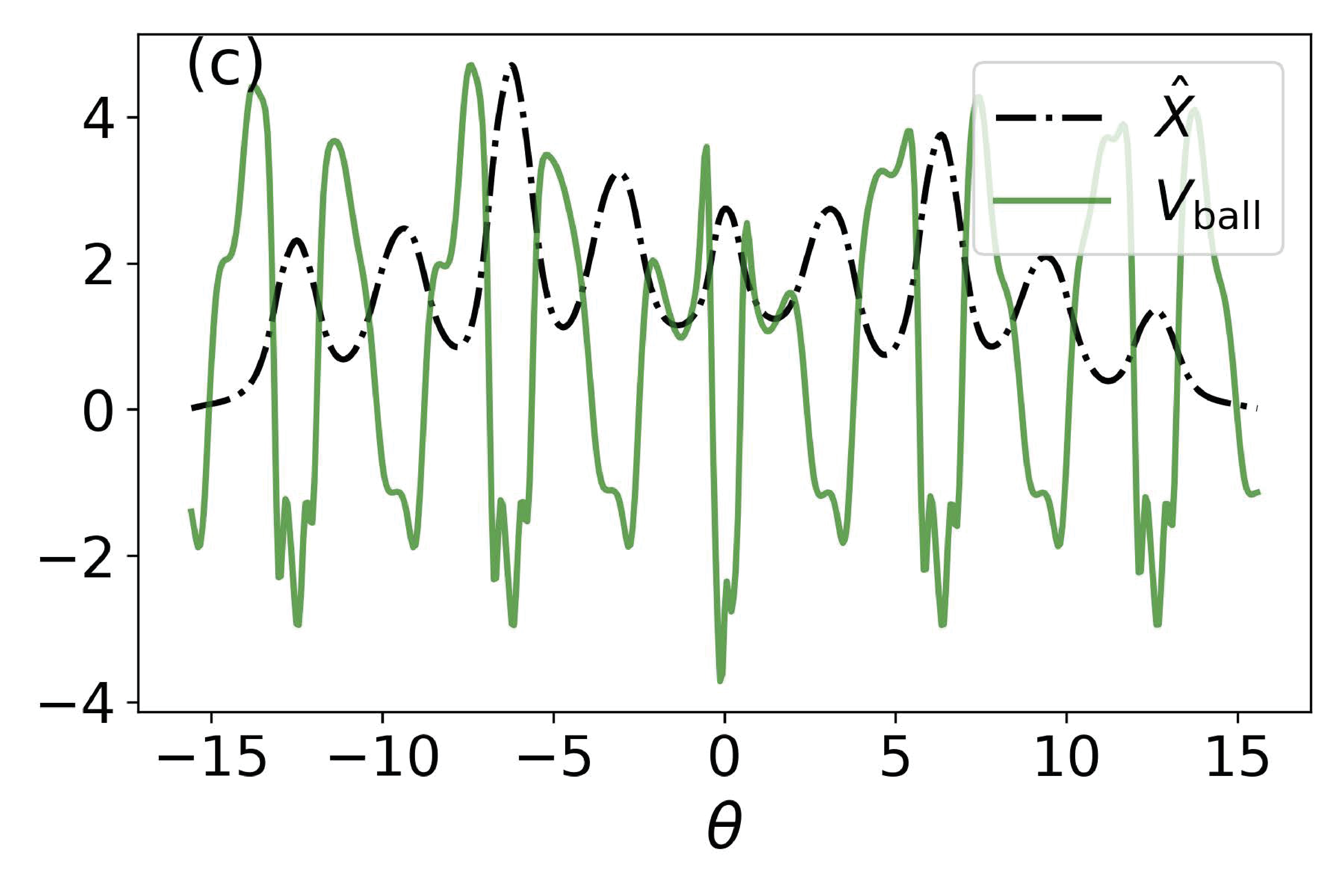}
    \includegraphics[height= 0.30\linewidth]{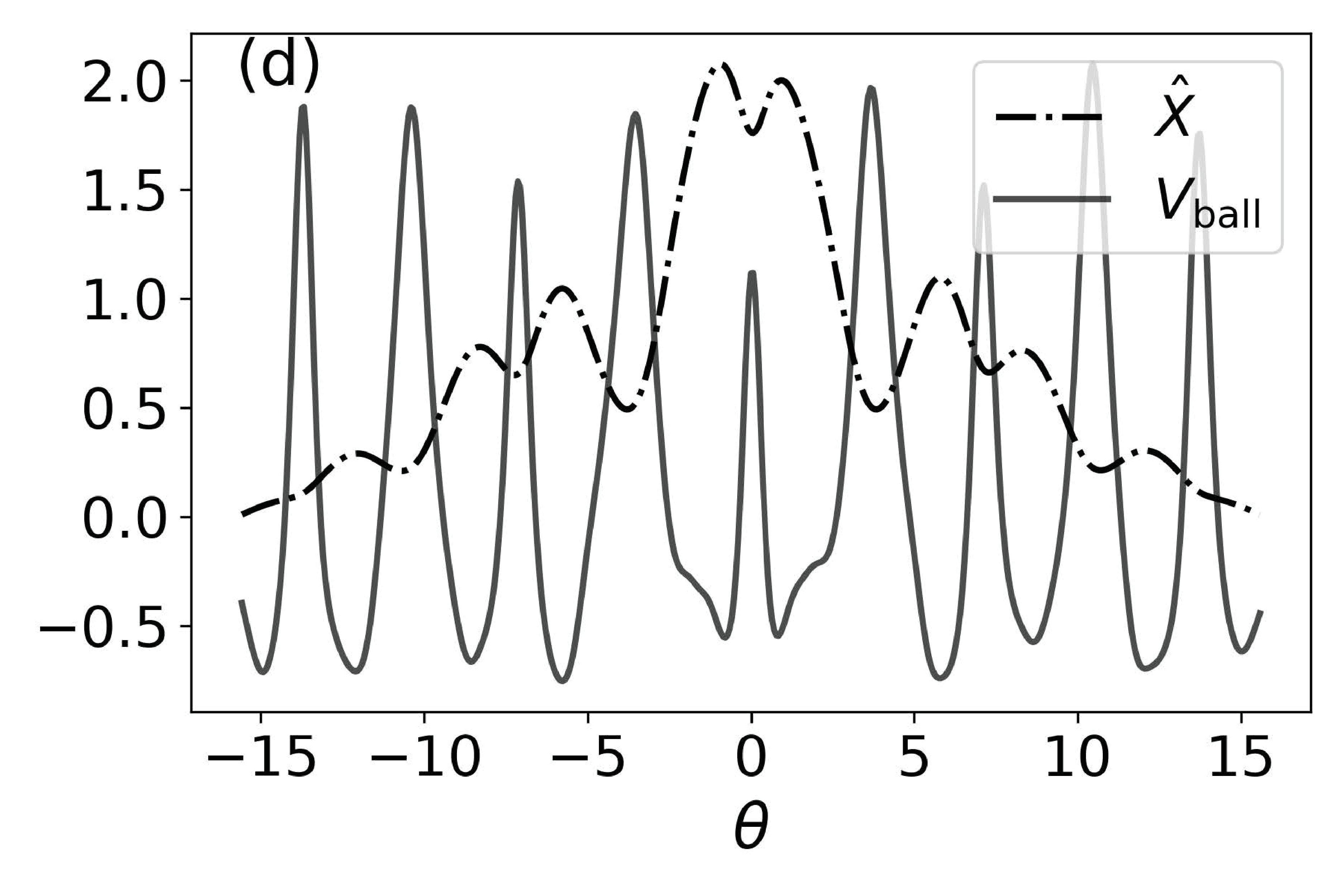} \\
    \caption{Eigenfunction $\hat{X}$ and effective ballooning potential at the maximum growth rate for the original and perturbed ITER equilibria. In each figure, the eigenfunctions have been scaled by a factor of the maximum effective ballooning potential. Figures $\mathrm{(a)}$ and $\mathrm{(c)}$ correspond to the aspect ratios of five stellarator and ITER cases, respectively. Similarly, $\mathrm{(b)}$ and $\mathrm{(d)}$ correspond to the aspect ratio of ten stellarator and ITER cases, respectively.}
    \label{fig:arbit_gamma_ball_AL}
\end{figure}

To gain insight into the ballooning stability of the ITER-like equilibria, we calculate the effective ballooning by applying the transformation $\hat{X} = \sqrt{g} X$ to~\eqref{eqn:ideal-ballooning-equation}. The transformed ballooning equation becomes
\begin{equation}
    \frac{d^2 \hat{X}}{d \theta^2} + \lbr E-V_{\mathrm{ball}}\rbr \hat{X} = 0, \quad E = -\frac{\lambda f}{g},
\end{equation}
where
\begin{equation}
        -V_{\mathrm{ball}} = \frac{c}{g} - \frac{d}{d\theta}\left(\frac{a}{2}\right) - \frac{a^2}{4}, \quad a = \frac{d\log(g)}{d \theta}.
\end{equation}
is called the effective ballooning potential. We plot the effective ballooning potential at the maximum growth rate values for the four growth rate plots from Figure~\ref{fig:arbit_gamma_ball} in Figure~\ref{fig:arbit_gamma_ball_AL} together with the normalized eigenfunction $\hat{X}$.
We can see that the stellarator equilibria potentials have higher peaks compared to tokamaks, which cause the ballooning eigenfunction to decay rapidly, similar to the decay of a wavefunction in a potential well in Schrodinger's equation. Towards the edge, we also observe more oscillations in stellarator potentials as seen in Figure~\ref{fig:arbit_gamma_ball_AL}$\mathrm{(a)}$ due to their three-dimensional shape. This decay of the eigenfunction is similar to the phenomenon of Anderson localization~\citep{anderson1958absence}. Multiple studies by~\cite{redi2002anderson,redi2001localized} have demonstrated the importance of Anderson localization of ballooning modes due to perturbations that break axisymmetry of the background equilibrium. Further detailed analysis of the localization of the 3D modes will be published elsewhere.

\section{Discussion and Conclusion}\label{sec:discussion}
In this work, we have shown how the overdetermined problem of volumetric quasisymmetric MHD equilibrium can be approximately solved by utilizing the inverse-aspect-ratio of a stellarator as an expansion parameter. Our assumptions, namely finite rotational transform, large aspect-ratio $\ep\ll 1$, and high plasma beta ($\beta\sim O(\ep)$), are consistent with the High-Beta Stellarator (HBS) model \citep{freidberg2014idealMHD}. With a purely toroidal vacuum field and a constant field strength as our lowest-order magnetic field, we show that the quasiaxisymmetric MHD equilibrium can be described by a Grad-Shafranov equation, much like axisymmetry. In particular, we show that approximate quasi-axisymmetric equilibria can be obtained from large-aspect-ratio axisymmetric Grad-Shafranov solutions by breaking axisymmetry through a periodic up-down deformation of the surfaces. Our study of Mercier and ballooning stability of some of these equilibria shows that the nonaxisymmetric deformations do not significantly degrade the MHD stability properties.

The original HBS model (\cite{freidberg2014idealMHD}) describes MHD equilibrium in large aspect-ratio stellarators. HBS consists of two coupled nonlinear PDEs: magnetic differential equations for the pressure surface and the parallel current. Being fully three-dimensional and nonlinear, they inherit the same problems as 3D MHD: the nonexistence of nested flux surfaces, the possibility of current singularities, and the formation of magnetic islands. As discussed in \citep{freidberg2014idealMHD}, the nonlinearity and three-dimensionality present a severe hindrance to constructing analytical equilibria. An exception is the Greene-Johnson \citep{Greene_Johnson_1961_Stallarator_expansion} limit of the HBS model, $N\sim \ep^{-1/2}\gg 1$, for a classical stellarator \citep{wakatani1998stellarator_heliotron,freidberg2014idealMHD,loizu2017equilibrium_beta,baillod2022equilibrium_beta_bootstrap}, where a GS equation can be obtained by an averaging over the short helical wavelength associated with large $N$. 

In this work, we have imposed volumetric quasisymmetry on the HBS model, allowing us to integrate the magnetic differential equation for the parallel current and obtain a 3D Grad-Shafranov equation for the flux surface label without imposing large $N$. The analysis of the fully 3D equilibria parallels that of an axisymmetric tokamak, thanks to the constraints from quasisymmetry, consistent with the large aspect-ratio limit of \citet{burby2020}. Since volumetric QS and MHD are not satisfied exactly but only to the lowest nontrivial order in $\ep$, our results are also consistent with \cite{plunk2019}, \cite{plunk2020_near_axisymmetry_MHD}, and the CDG Theorem \citep{constantin2021}.

Our work was motivated by the success of the second-order near-axis expansions in providing an analytic description of quasisymmetric stellarators. While the near-axis approach works for any stellarator, provided we zoom in sufficiently close to the magnetic axis, the present approach requires the stellarator to have a large aspect ratio, which is quite reasonable for most stellarators. A significant advantage of our approach over the near-axis expansion is a global description of MHD equilibrium free from any polynomial expansion in the radial variable. Therefore, magnetic shear, currents, and pressure profiles can be quite general. Furthermore, we can allow any flux surface shaping. In contrast, the second-order near-axis expansion only allows triangularity in shaping. 

The quasisymmetric Grad-Shafranov equation admits several classes of exact analytical solutions. In fact, any analytic solutions of the large-aspect-ratio axisymmetric Grad-Shafranov equation can generate a quasisymmetric solution through a straightforward coordinate transform that induces a periodic up-down deformation. Utilizing this, we have generated numerical solutions starting from an axisymmetric tokamak equilibrium and deforming it accordingly. We have then verified the validity of this methodology using the VMEC code. As the aspect ratio gets larger, the deviations from the periodically up-down shifted tokamak equilibrium, and the actual stellarator equilibrium generated with the same boundary using the VMEC code becomes smaller.

There are several directions in which the current work can be extended. First, our current model can only describe quasiaxisymmetric configurations close to axisymmetry due to the choice of the lowest-order vacuum field being a purely toroidal field. A more general choice for the lowest-order vacuum field is needed to describe quasiaxisymmetry and quasihelical symmetry. Second, the isomorphism between QS-HBS and axisymmetric tokamak equilibrium suggests that we can carry out a local geometry analysis for QS-HBS in close analogy to the axisymmetric Miller-geometry case \citep{miller_geometry1998noncircular}. Near-axisymmetric local equilibria could then be used to study kinetic ballooning modes, ITG, and ETG thresholds. Third, one can study subsonic flows in quasisymmetric stellarators, which have been predicted \citep{spong2005flow_generation} and experimentally demonstrated \citep{gerhardt2005HSX_flows_PRL}, using the QS-HBS model. The QS-HBS model should offer interesting analytical insights into the structure of subsonic flows and their damping. Fourth, the HBS model in the Greene-Johnson limit has been used to study the plasma $\beta$-limit in a stellarator, which is of practical importance \citep{freidberg2014idealMHD,loizu2017equilibrium_beta,baillod2022equilibrium_beta_bootstrap}. An analogous study can be conducted for the QS-HBS model to understand how QS impacts the $\beta$ limit. Finally, we hope that the near-axisymmetric aspect of the QS-HBS model can provide an avenue to study nonaxisymmetric deformations of tokamaks that preserved quasisymmetry and hence can be successfully used in controlling tokamaks using 3D perturbations \citep{Park_QA}.

\noindent{\bf Acknowledgements}
The authors would like to thank F. P. Diaz, V. Duarte, H. Weitzner, E. J. Paul, M. Zarnstroff, J. Loizu, A. Baillod, H. Zhu, R. Jorge, E. Rodr\'iguez, and M. Landreman for stimulating discussions and helpful suggestions.

This research was supported by a grant from the Simons Foundation/SFARI (560651, AB), and the Department of Energy Award No. DE-SC0024548.

\section{Appendices}
\appendix

\section{Consistency of QS-GSE and $\dpl \Psi=0$}\label{app:consistency}
We shall now investigate the consistency of the following set of equations:
\begin{subequations}
    \begin{align}
     \dlrs \Psi+ 2 x\frac{d\beta(\Psi)}{d\Psi} + H'(\Psi)-\mfa_{,xx}=0, \quad \dlrs = \del_x^2+\del_y^2 \label{eq:QSGSE_app},
     \\
     \cL \Psi=0, \quad \cL=  \lbr \del_\phi - \mfa_{,x}\del_y\rbr \label{eq:dpl_Psi_cL_op},
\end{align}
\label{eq:consistency_set}
\end{subequations}
obtained from \eqref{eq:QSGSE} and \eqref{eq:uDpsi}. The consistency condition follows from the fact that $\cL$ on equation \eqref{eq:QSGSE_app} should vanish identically. We shall use the following identities, which follow from straightforward, direct calculations using \eqref{eq:dpl_Psi_cL_op}:
\begin{subequations}
    \begin{align}
    \cL \dlrs \Psi= \lbr \del_x^3 \mfa\rbr \del_y \Psi +2 \lbr \del_x^2 \mfa \rbr\del^2_{xy}\Psi,\\
    \cL \lbr  2 x\frac{d\beta(\Psi)}{d\Psi} + H'(\Psi) \rbr =0,\\
    \cL \mfa_{,xx}= \del_\phi (-\mfa_{,xx}).
    \end{align}
    \label{eq:some_ids}
\end{subequations}
Using the above identities, we can rewrite the consistency condition as an equation for $\mfa$ 
\begin{align}
    \lbr \Psi_{,y} \del_x +2 \Psi_{,xy}-\del_\phi\rbr \zeta=0, \quad \zeta=\mfa_{,xx}.
    \label{eq:consistency_cond_mfa}
\end{align}
Using the fact that $\mfa$ is $y-$independent, we can rewrite \eqref{eq:consistency_cond_mfa} as
\begin{align}
    \del^2_{x y} \lbr \sqrt{\zeta} \Psi\rbr - \del_\phi \sqrt{\zeta}=0.
    \label{eq:sqrt_zeta_eq}
\end{align}
Averaging over $\phi$ and using periodicity of $\mfa, \zeta$ in $\phi$ we obtain
\begin{align}
   \del^2_{xy} \oint d\phi \sqrt{\zeta}\;\Psi =0 \quad \Leftrightarrow \quad \oint d\phi \sqrt{\zeta}\;\Psi = \text{constant}.
   \label{eq:ortho_sqrt_zeta}
\end{align}
We note that $\Psi$ depends on all three variables whereas $\mfa, \zeta$ depends only on $(x,\phi)$. Thus, satisfying \eqref{eq:consistency_cond_mfa} or the constraint \eqref{eq:ortho_sqrt_zeta}, with a nonzero $\zeta$ is in general not possible. Thus, a possible solution is $\mfa_{,xx}=0$, which implies that $\mfa$ is linear in $x$. Thus, we obtain \eqref{eq:axx_is_0}.

\section{Comparison with Burby-Kallinikos-MacKay}\label{app:BKM}
\cite{burby2020} obtained the following generalized quasisymmetric Grad-Shafranov equation: 
\begin{align}
    \dl^2 \psi - \frac{\u \times \v}{u^2}\cdot \dl \psi + \frac{\u\cdot \v}{u^2}\u\cdot \B -\u \cdot \J =0,
    \label{eq:Burby_preGS}
\end{align}
where, $\u$ is the quasisymmetry vector, $\v=\dl \times \u$. Note that in \eqref{eq:Burby_preGS}, $\psi$ denotes $\bm{A}\cdot \u$ with $\bm{A}$ denoting the vector potential. For MHD equilibrium, $\u\cdot \B =F(\psi)$ and \eqref{eq:Burby_preGS} takes the form
\begin{align}
     \dl^2 \psi - \frac{\u \times \v}{u^2}\cdot \dl \psi + \frac{\u\cdot \v}{u^2}F +FF'-u^2 p'(\psi)=0.
    \label{eq:Burby_QGS}
\end{align}
For axisymmetry and helical symmetry, we have, respectively
    \begin{align}
    \u&=R^2\dl\phi,\; u=R,\; \v= -2 \bm{e_Z},\; \u\cdot \v=0,\; \u\times \v=2 R \bm{e_R}, \qquad\qquad\qquad\quad\;\\
    \u&=R^2\dl\phi-l \dl Z,\; u=\sqrt{R^2+l^2},\; \v=-2 \bm{e_Z},\; \u \cdot \v = 2 l,\;  \u\times \v=2 R \bm{e_R}.\nonumber
\end{align}

Equation \eqref{eq:Burby_QGS} reduces to the standard axisymmetric and helically symmetric Grad-Shafranov equations, respectively.

We now derive the expression for $\u$. We use the following identities, obtained in Sections \ref{sec:QSHBS_derivation} and \ref{secQS_GS_derivation}:
\begin{align}
\frac{d\Psi}{d\psi}=\frac{1}{\ep F}, \quad \frac{F}{B_0}=\frac{1}{\ep \iota}(1-\ep \beta), \quad \frac{B}{B_0}=1-\ep(x+\beta), \quad \dpl x = -\del_\xi\Psi(x,\xi).
    \label{eq:ids_for_BKM}
\end{align}
Starting with the definition for $\u$ \eqref{eq:QS_u}, we find that
\begin{align}
    \u = \frac{\dl \Psi \times \dl B}{B \lbr \ep \dpl B\rbr}\lbr \frac{d\Psi}{d\psi} \rbr^{-1}= \frac{F/B_0}{B/B_0} \frac{\dl \Psi \times \dl x}{\dpl x}.
\end{align}
Using the identities \eqref{eq:ids_for_BKM} together with the definition of $\xi$ \eqref{eq:trav_wave_Psi}, $\u$ takes the form
\begin{align}
    \u=\frac{1}{\ep \iota}(1-\ep \beta)(1+\ep (x+\beta))\frac{\dl \xi\times \dl x}{\dpl x}\Psi_{,\xi} = \frac{1}{\ep \iota}\lbr \bm{e_\phi}(1+\ep x)-\ep \mfa_{,x}\bm{e_Z}\rbr.
    \label{eq:derivation_u}
\end{align}
The expression for $\u_{\text{HBS}}=\iota \u$ is
\begin{align}
    \u_{\text{HBS}}= R^2 \dl \phi -\mfa_{,x}\dl Z = \bm{e_\phi}\lbr \frac{1}{\ep}+x\rbr -\mfa_{,x}\bm{e_Z},
\end{align}
which follows from \eqref{eq:derivation_u} and the definition $R=\ep^{-1}+x$. 

We now show that \eqref{eq:Burby_QGS} for HBS takes the same form as the QS-GSE \eqref{eq:QSGSE_TW}. Substituting the following identities into \eqref{eq:Burby_QGS}
\begin{subequations}
    \begin{align}
|\u_{\text{HBS}}|=\frac{1}{\ep}+x,\; \v_{\text{HBS}}=-2 \bm{e_Z}-\ep \mfa_{x\phi}\bm{e_R},\; FF'=-\frac{1}{\ep}\beta' +O(1) \\
\frac{\u_{\text{HBS}}\cdot \v_{\text{HBS}}}{|\u_{\text{HBS}}|^2} = -2\mfa_{,x} \ep^2, \quad \frac{\u_{\text{HBS}}\times \v_{\text{HBS}}}{|\u_{\text{HBS}}|^2} = 2 \ep \bm{e_R},
    \label{eq:uvw_xpressions}
\end{align}
\end{subequations}
we find that both the $\u\times \v$ and the $\u \cdot \v$ terms in \eqref{eq:Burby_QGS} are negligible to the lowest order in $\ep$. Also, the $FF'$ and the pressure term cancel to the lowest order, leaving the $2x\beta'$ term in \eqref{eq:QSGSE_TW}. As a result, we get \eqref{eq:QSGSE_TW} with the identification that $H'(\Psi)$ is the $O(1)$ term of the expression $FF'$. 

Finally, let us show that $\u_{\text{HBS}}$ is not a Killing vector. As discussed in \citep{burby2020}, a vector $\u$ is Killing if, and only if, the vector $\bm{w}$ is identically zero, where
\begin{align}
    \w= -\u \times \v + \dl u^2, \quad \v =\dl \times \u.
\end{align}
Evaluating $\w_{\text{HBS}}$ we find it to be non-zero since
\begin{align}
    \w_{\text{HBS}}= -\mfa_{,x\phi} \bm{e_Z}= Y''(\phi) \bm{e_Z}\neq 0
\end{align}

\section{Various coordinate transformations}
\label{app:coords}
Navigating between different coordinate systems is essential for simplifying PDEs and obtaining solutions. In this work, we find the following coordinate systems, besides the HBS coordinates $(x,y,\phi)$, to be advantageous:
\begin{align}
(x,\xi,\phi), \quad (\Psi,\xi,\phi), \quad (\Psi,\alpha,\phi), \quad \text{and}\quad (\Psi,\alpha,\ell).
    \label{eq:various_coords}
\end{align}
We begin with the transformation from $(x,y,\phi)$ to $(x,\xi= y +Y(\phi),\phi)$,
\begin{align}
    \at{\del_x}{(y,\phi)}=\at{\del_x}{(\xi,\phi)}, \quad \at{\del_y}{(x,y)}=\at{\del_\xi}{(x,\phi)}, \quad \at{\del_\phi}{(x,y)}= \at{\del_\phi}{(x,\xi)}+Y'(\phi)\del_\xi.
    \label{eq:xy_to_x_xi}
\end{align}
The $\dpl$ operator in $(x,y,\phi)$ is
\begin{align}
    \dpl = \del_\phi -Y'(\phi)\del_y +\{\Psi,\;\;\}_{(x,y)},
    \label{eq:dpl_in_xy}
\end{align}
which follows from  $\mfa_{,x}=Y'(\phi)$, and the definition of $\dpl$ given in \eqref{eq:dpl_dlrs_def}. Using \eqref{eq:xy_to_x_xi}, the expression for $\dpl$ can be rewritten as
\begin{align}
    \dpl = \at{\del_\phi}{(x,\xi)}+\{\Psi,\;\;\}_{(x,\xi)}.
    \label{eq:dpl_in_x_xi}
\end{align}
Thus, $\dpl\Psi= \del_\phi \Psi=0$ in the $(x,\xi,\phi)$ coordinates, pointing to the 2D nature of $\Psi$.

We now exploit the two-dimensionality of $\Psi$ to transform to the $(\Psi,\xi,\phi)$ coordinates using
\begin{align}
     \at{\del_x}{(y,\phi)}=\Psi_{,x}\at{\del_\Psi}{(\xi,\phi)}, \quad \at{\del_\xi}{(x,y)}=\at{\del_\xi}{(x,y)}+\Psi_{,\xi}\at{\del_\Psi}{(\xi,\phi)}, \quad \at{\del_\phi}{(x,\xi)}= \at{\del_\phi}{(\Psi,\xi)}.
    \label{eq:x_xi_to_Psi_xi}
\end{align}
The parallel derivative $\dpl$ \eqref{eq:dpl_in_x_xi}, now takes the form
\begin{align}
    \dpl= \at{\del_\phi}{(\Psi,\xi)} +\Psi_{,x}\at{\del_\xi}{(\Psi,\phi)}.
    \label{eq:dpl_in_Psi_xi}
\end{align}
It follows from the fact that $\Psi_{,x}$ is only a function of $\Psi,\xi$, $\dpl \alpha=0$ and $\ep\dpl\ell =1$ that
\begin{align}
    \alpha = \phi - \int_\Psi \frac{d\xi}{\Psi_{,x}},\quad \ep \ell = \phi +L(\Psi).
    \label{eq:alpha_and_ell_forms}
\end{align}
To check the validity of \eqref{eq:alpha_and_ell_forms}, we now check that we can recover $\B$ as given in \eqref{eq:B_form} from $\dl\alpha\times\dl\Psi$. From \eqref{eq:alpha_and_ell_forms}, we have
\begin{align}
    \dl \alpha \times \dl \Psi = -\dl \Psi \times \dl \phi +\dl \Psi \times \dl \xi \frac{1}{\Psi_{,x}}.
    \label{eq:pre_B_Clebsh}
\end{align}
We can simplify the last term in \eqref{eq:pre_B_Clebsh} in $(x,y,\phi)$ coordinates using $\Psi=\Psi(x,\xi),\xi=y+Y(\phi), \mfa_{,x}=Y'(\phi)$. We obtain 
\begin{align}
    \dl \Psi \times \dl \xi \frac{1}{\Psi_{,x}}&= \Psi_{,x}\dl x \times \dl (y+Y(\phi)) \frac{1}{\Psi_{,x}},\\
    &= \bm{e_\phi}+Y'(\phi)\dl x \times \dl \phi\nonumber,\\
    &= \bm{e_\phi}+\dl \mfa \times \dl \phi\nonumber.
\end{align}
Therefore,
\begin{align}
    \dl \alpha \times \dl \Psi =\bm{e_\phi}+\dl (\mfa-\Psi) \times \dl \phi = \bm{e_\phi}+\dlr \cA \times \dl \phi,
    \label{eq:B_Clebsh}
\end{align}
upon using $\Psi+\cA=\mfa$.

Comparing with the expression of $\B$ from  \eqref{eq:B_form}, we find that the $\bm{e_\phi}(x+\beta)$ terms are mixing. Including the $O(\ep)$ correction to $\alpha$ such that
\begin{align}
    \alpha = \phi - \int_\Psi d\xi \frac{B_\phi/B_0}{\Psi_{,x}},
\end{align}
remits this problem. We need not just $\alpha$ but also $\dl \alpha$ to be accurate to $O(\ep)$. The first term in $\alpha$, i.e., $\phi$, has a gradient of $O(\ep)$. Thus, we also need the $O(\ep)$ piece from $B_\phi/B_0$ to contribute to $\dl \alpha$ to $O(\ep)$. 

\section{On-axis rotational transform from the integrated torsion}
The axis of the QS-HBS system is a circle deformed periodically in the $Z$ direction. Let us now obtain an expression for total torsion ($\oint \tau d\ell$), representing the torsion contribution to the on-axis rotational transform formula due to Mercier.

Using cylindrical coordinates $(R, Z, \phi)$, we can denote the axis by the following space curve,
\begin{align}
    \bm{r}=R_0 (\bm{e_R}+\mff(\phi)\bm{e_Z}).
\end{align}
Here, the first term represents a circle of constant radius $R_0$, and $\mff$ is a periodic deformation of the same. Using standard relations
\begin{align}
    \del_\phi \bm{e_R}=\bm{e_\phi} , \quad  \del_\phi \bm{e_\phi}=-\bm{e_R},
\end{align}
we find the arclength $\ell$, and the unit tangent vector to be
\begin{align}
    \frac{d\ell}{d\phi}= R_0\sqrt{1+\mff'(\phi)^2}, \quad \bm{t}\equiv \frac{d\bm{r}}{d\ell} = \frac{(\bm{e_\phi}+\mff' \bm{e_Z})}{\sqrt{1+\mff'^2}}.
\end{align}
We shall use the following identities for the curvature and torsion, $\kappa,\tau$,
\begin{align}
    \kappa= \bigg|\frac{d\bm{r}}{d\ell}\times \frac{d^2\bm{r}}{d\ell^2}\bigg|, \quad \tau = \frac{1}{\kappa^2}\frac{d\bm{r}}{d\ell}\times \frac{d^2\bm{r}}{d\ell^2}\cdot \frac{d^3\bm{r}}{d\ell^3}.
\end{align}
Straightforward calculation of the higher derivatives of $\bm{r}$ leads to the following expressions for $\kappa,\tau$:
\begin{align}
    \kappa = \frac{1}{R_0} \sqrt{\frac{1+\mff'^2 +\mff''^2}{(1+\mff'^2)^3}}, \quad \tau = \frac{1}{R_0} \frac{\mff' +\mff'''}{1+\mff'^2 +\mff''^2}.
\end{align}
Therefore,
\begin{align}
    \oint \tau d\ell = \oint d\phi \sqrt{1+\mff'^2}\frac{\mff' +\mff'''}{1+\mff'^2 +\mff''^2}.
    \label{eq:tau_dl_gen_exp}
\end{align}
In general, this is a non-vanishing quantity. However, since the deformation $\mff$ can be only as large as the minor radius, i.e., $\mff= \ep \mff_1$, the integrated torsion is not generally an $O(1)$ quantity as we now show.

We find that
\begin{align}
    \oint \tau d\ell = \ep \oint d\phi (\mff_1' +\mff_1''') + \ep^3 \oint d\phi (\mff_1' +\mff_1''')(\mff''_1)^2 +O(\ep^5).
\end{align}
The first term in the expression for torsion averages out identically. The next term is $O(\ep^3)$ if the toroidal harmonic, $n\sim 1$. However, for sizable $n$, we can get an amplification factor coming from the derivatives of $\mff$. To estimate this factor, we note that the total torsion can be further simplified to 
\begin{align}
   \oint \tau d\ell = \ep^3 \oint d\phi\; \mff'_1 \mff_1''^2,
   \label{eq:total_torsion_exp}
\end{align}
since the $\mff_1'''$ term is a total derivative.

We observe several facts from \eqref{eq:total_torsion_exp}. First, if $\mff_1(\phi)$ is even or simple-harmonic, i.e., $\exp(i n \phi)$, the total torsion vanishes. The reason why a simple-harmonic $\mff_1$ leads to a zero total torsion is that, in this case, we can always add a term $n^2 \mff_1''' f_1''^2$, which vanishes by itself. However, this term cancels the $\mff_1'f_1''^2$ term due to the simple-harmonic nature of $\mff_1$. So, consider a two-harmonic stellarator-symmetric deformation: $\mff_1 = a\sin{n\phi} + b\sin{m\phi}$. After some straightforward algebra, one gets:
\begin{equation}
\begin{aligned}
    \oint\mff_1'\mff_1''d\phi = \int_0^{2\pi} \Bigg[&a^2 bmn\left(mn^2\sin{m\phi}\sin{2n\phi} - \frac{n^3}{2}\cos{m\phi}\cos{2n\phi}\right) \\
    &+ ab^2 mn\left(m^2 n\sin{n\phi}\sin{2m\phi} - \frac{m^3}{2}\cos{n\phi}\cos{2m\phi}\right)\Bigg]d\phi.
\end{aligned}
\end{equation}
Clearly, in order for the integral to be nonzero, it must be that either $m=2n$ or $n=2m$. Without loss of generality, take $m=2n$. Then,
\begin{equation}
    \iota_\tau = \frac{1}{2\pi}\oint\tau d\ell = \frac{\ep^3}{2\pi}\oint\mff_1'\mff_1''d\phi = \frac{3}{2}\ep^3 a^2 bn^5.
    \label{eq:iota_tau}
\end{equation}
We note the strong amplification factor in the above expression for the rotational transform due to the $n^5$ scaling. However, note that if the ordering $\partial_\phi \sim \ep\nabla_\perp$ is to be valid, one must have $a\sim 1/n$ and $b\sim 1/m$; and thus $\iota_\tau = O(\ep^3 n^2)$. This implies that low-amplitude large-$n$ modes could have a sizable torsional contribution to $\iota$. However, very large $n$ would lead to higher derivatives growing faster, leading to the breakdown of the LAE. Hence, $n\leq 3$ is perhaps the best choice for $n$, which limits the torsional contribution to $\iota_\tau \sim 0.05$ for an aspect ratio of $\sim 5$.


\bibliographystyle{jpp}
\bibliography{plasmalit}

\end{document}